\documentstyle[12pt]{article}

\advance \topmargin by -\headheight
\advance \topmargin by -\headsep
     
\evensidemargin \oddsidemargin
\begin{document}
\newcommand{\sect}[1]{\setcounter{equation}{0}\section{#1}}
\renewcommand{\theequation}{\thesection.\arabic{equation}}

\topmargin -.6in
\def\nonu{\nonumber}
\def\rf#1{(\ref{eq:#1})}
\def\lab#1{\label{eq:#1}} 
\def\br{\begin{eqnarray}}
\def\er{\end{eqnarray}}
\def\be{\begin{equation}}
\def\ee{\end{equation}}
\def\0{\nonumber}
\def\lb{\lbrack}
\def\rb{\rbrack}
\def\({\left(}
\def\){\right)}
\def\v{\vert}
\def\bv{\bigm\vert}
\def\lskip{\vskip\baselineskip\vskip-\parskip\noindent}
\relax
\newcommand{\nit}{\noindent}
\newcommand{\ct}[1]{\cite{#1}}
\newcommand{\bi}[1]{\bibitem{#1}}
\def\a{\alpha}
\def\b{\beta}
\def\ca{{\cal A}}
\def\cm{{\cal M}}
\def\cn{{\cal N}}
\def\cf{{\cal F}}
\def\d{\delta}
\def\D{\Delta}
\def\eps{\epsilon}
\def\g{\gamma}
\def\G{\Gamma}
\def\grad{\nabla}
\def\h{ {1\over 2}  }
\def\hc{\hat{c}}
\def\hd{\hat{d}}
\def\hg{\hat{g}}
\def\hp{ {+{1\over 2}}  }
\def\hm{ {-{1\over 2}}  }
\def\k{\kappa}
\def\l{\lambda}
\def\L{\Lambda}
\def\lg{\langle}
\def\m{\mu}
\def\n{\nu}
\def\o{\over}
\def\om{\omega}
\def\O{\Omega}
\def\p{\phi}
\def\pa{\partial}
\def\pr{\prime}
\def\ra{\rightarrow}
\def\rh{\rho}
\def\rg{\rangle}
\def\s{\sigma}
\def\t{\tau}
\def\th{\theta}
\def\ti{\tilde}
\def\wti{\widetilde}
\def\inte{\int dx }
\def\xb{\bar{x}}
\def\yb{\bar{y}}

\def\tr{\mathop{\rm tr}}
\def\Tr{\mathop{\rm Tr}}
\def\partder#1#2{{\partial #1\over\partial #2}}
\def\ds{{\cal D}_s}
\def\wtwo{{\wti W}_2}
\def\lie{{\cal G}}
\def\alie{{\widehat \lie}}
\def\dlie{{\cal G}^{\ast}}
\def\elie{{\widetilde \lie}}
\def\edlie{{\elie}^{\ast}}
\def\hlie{{\cal H}}
\def\wlie{{\widetilde \lie}}

\def\rlx{\relax\leavevmode}
\def\inbar{\vrule height1.5ex width.4pt depth0pt}
\def\IZ{\rlx\hbox{\sf Z\kern-.4em Z}}
\def\IR{\rlx\hbox{\rm I\kern-.18em R}}
\def\IC{\rlx\hbox{\,$\inbar\kern-.3em{\rm C}$}}
\def\one{\hbox{{1}\kern-.25em\hbox{l}}}

\def\PRL#1#2#3{{\sl Phys. Rev. Lett.} {\bf#1} (#2) #3}
\def\NPB#1#2#3{{\sl Nucl. Phys.} {\bf B#1} (#2) #3}
\def\NPBFS#1#2#3#4{{\sl Nucl. Phys.} {\bf B#2} [FS#1] (#3) #4}
\def\CMP#1#2#3{{\sl Commun. Math. Phys.} {\bf #1} (#2) #3}
\def\PRD#1#2#3{{\sl Phys. Rev.} {\bf D#1} (#2) #3}
\def\PRB#1#2#3{{\sl Phys. Rev.} {\bf B#1} (#2) #3}

\def\PLA#1#2#3{{\sl Phys. Lett.} {\bf #1A} (#2) #3}
\def\PLB#1#2#3{{\sl Phys. Lett.} {\bf #1B} (#2) #3}
\def\JMP#1#2#3{{\sl J. Math. Phys.} {\bf #1} (#2) #3}
\def\PTP#1#2#3{{\sl Prog. Theor. Phys.} {\bf #1} (#2) #3}
\def\SPTP#1#2#3{{\sl Suppl. Prog. Theor. Phys.} {\bf #1} (#2) #3}
\def\AoP#1#2#3{{\sl Ann. of Phys.} {\bf #1} (#2) #3}
\def\PNAS#1#2#3{{\sl Proc. Natl. Acad. Sci. USA} {\bf #1} (#2) #3}
\def\RMP#1#2#3{{\sl Rev. Mod. Phys.} {\bf #1} (#2) #3}
\def\PR#1#2#3{{\sl Phys. Reports} {\bf #1} (#2) #3}
\def\AoM#1#2#3{{\sl Ann. of Math.} {\bf #1} (#2) #3}
\def\UMN#1#2#3{{\sl Usp. Mat. Nauk} {\bf #1} (#2) #3}
\def\FAP#1#2#3{{\sl Funkt. Anal. Prilozheniya} {\bf #1} (#2) #3}
\def\FAaIA#1#2#3{{\sl Functional Analysis and Its Application} {\bf #1} (#2)
#3}
\def\BAMS#1#2#3{{\sl Bull. Am. Math. Soc.} {\bf #1} (#2) #3}
\def\TAMS#1#2#3{{\sl Trans. Am. Math. Soc.} {\bf #1} (#2) #3}
\def\InvM#1#2#3{{\sl Invent. Math.} {\bf #1} (#2) #3}
\def\LMP#1#2#3{{\sl Letters in Math. Phys.} {\bf #1} (#2) #3}
\def\IJMPA#1#2#3{{\sl Int. J. Mod. Phys.} {\bf A#1} (#2) #3}
\def\AdM#1#2#3{{\sl Advances in Math.} {\bf #1} (#2) #3}
\def\RMaP#1#2#3{{\sl Reports on Math. Phys.} {\bf #1} (#2) #3}
\def\IJM#1#2#3{{\sl Ill. J. Math.} {\bf #1} (#2) #3}
\def\APP#1#2#3{{\sl Acta Phys. Polon.} {\bf #1} (#2) #3}
\def\TMP#1#2#3{{\sl Theor. Mat. Phys.} {\bf #1} (#2) #3}
\def\JPA#1#2#3{{\sl J. Physics} {\bf A#1} (#2) #3}
\def\JSM#1#2#3{{\sl J. Soviet Math.} {\bf #1} (#2) #3}
\def\MPLA#1#2#3{{\sl Mod. Phys. Lett.} {\bf A#1} (#2) #3}
\def\JETP#1#2#3{{\sl Sov. Phys. JETP} {\bf #1} (#2) #3}
\def\JETPL#1#2#3{{\sl  Sov. Phys. JETP Lett.} {\bf #1} (#2) #3}
\def\PHSA#1#2#3{{\sl Physica} {\bf A#1} (#2) #3}
\def\PHSD#1#2#3{{\sl Physica} {\bf D#1} (#2) #3}

\newcommand\twomat[4]{\left(\begin{array}{cc}  
{#1} & {#2} \\ {#3} & {#4} \end{array} \right)}
\newcommand\twocol[2]{\left(\begin{array}{cc}  
{#1} \\ {#2} \end{array} \right)}
\newcommand\twovec[2]{\left(\begin{array}{cc}  
{#1} & {#2} \end{array} \right)}

\newcommand\threemat[9]{\left(\begin{array}{ccc}  
{#1} & {#2} & {#3}\\ {#4} & {#5} & {#6}\\ {#7} & {#8} & {#9} \end{array} \right)}
\newcommand\threecol[3]{\left(\begin{array}{ccc}  
{#1} \\ {#2} \\ {#3}\end{array} \right)}
\newcommand\threevec[3]{\left(\begin{array}{ccc}  
{#1} & {#2} & {#3}\end{array} \right)}

\newcommand\fourcol[4]{\left(\begin{array}{cccc}  
{#1} \\ {#2} \\ {#3} \\ {#4} \end{array} \right)}
\newcommand\fourvec[4]{\left(\begin{array}{cccc}  
{#1} & {#2} & {#3} & {#4} \end{array} \right)}

\begin{titlepage}
\vspace*{-2 cm}
\noindent
\begin{flushright}
\end{flushright}

\vskip 1 cm
\begin{center}
{\Large\bf Solitons with Isospin  } \vglue 1  true cm
{ J.F. Gomes}$^{1}$,
 { G.M. Sotkov}$^{1,2}$ and { A.H. Zimerman}$^{1}$\\

\vspace{1 cm}

${}^1\;${\footnotesize Instituto de F\'\i sica Te\'orica - IFT/UNESP\\
Rua Pamplona 145\\
01405-900, S\~ao Paulo - SP, Brazil}\\
\vskip 1 cm
\noindent
${}^2\;${\footnotesize Institute for Nuclear Research  and Nuclear Energy,\\
Bulgarian Academy of Sciences,\\
 Tsarigradsko Chausse 72,\\
  BG-1784, Sofia, Bulgaria} \\

\vspace{1 cm}

\end{center}

\normalsize
\vskip 0.5cm

\begin{center}
{ {\bf ABSTRACT}}\\
\end{center}
\noindent
We study the symmetries of the soliton spectrum of a pair of T-dual integrable models, invariant under  
global $SL(2)_q\otimes U(1)$ transformations. They represent an integrable
perturbation of the reduced Gepner parafermions, based on certain gauged $SL(3)$ - WZW model.  
Their (semiclassical) topological soliton solutions, carrying isospin
and belonging  to the root of unity representations of $q$-deformed $SU(2)_q$ - algebra 
are obtained. We derive the semiclassical particle spectrum
of these models, which is  further used to  prove  their T-duality properties.
 
\end{titlepage}
  
\sect{Introduction}

Integrable perturbations of two-dimensional (2-D) conformal field theories (CFT's) are known to 
describe the off-critical behaviour
of a large class of 2-D statistical models near to
their second order phase transition \cite{z1} (see \cite{muss} for review). The corresponding 
massive integrable models (IM's) 
have been  identified with the abelian affine Toda
field theories (ATFT) at some specific value of their coupling constant \cite{cm}.
Few significant examples are given by (i) $E_l^{(1)}$- ATFT's  for $l=6,7,8$ \cite{fz}, \cite{sz},
 (ii) Sine-Gordon model as $\phi_{13}$ perturbations of the Virasoro minimal models (mms) \cite{yang}
(iii) $A_n^{(1)}$ - ATFT's as perturbations of $W_{n+1}$- algebra  mms \cite{vega}. 
The Lagrangian description of certain integrable
deformations of $Z_n$ -parafermionic models (PF) \cite{fat1}, of the WZW models 
\cite{braz},
of Gepner 
PF's \cite{gepner} and of the $V_{n+1}$- algebra mms \cite{annals} requires however another class 
of IM's known as non-abelian affine Toda field theories (NA-ATFT) \cite{lez-sav},\cite{laf}. 
 The two simplest examples 
of such IM's are  the complex sine-Gordon  model \cite{lund} (and its 
generalizations \cite{pousa}) and the $A_2^{(1)}$-dyonic IM describing
integrable perturbations of the $W_3^{(2)}$ and $V_3$ - algebra mm's \cite{annals}, \cite{elek}.  
Although both the abelian and non-abelian ATFT's can be realized as integrable deformations
of the gauged $G/H$ - WZW models we have to mention  few important
differences between them, encoded in the different gradations of the underlining   affine algebra.
The first is that the target space metric $g^{kl}(\varphi_m)$  in the kinetic term of their Lagrangians,  
i.e. $g^{kl}(\varphi_m)\pa \varphi_k \bar \pa \varphi_l$
is flat for the abelian models and quite nontrivial function of the 
fields $\varphi_m$ for the non-abelian ones. The second difference
concerns their symmetries:  the abelian ATFT's do not have Noether 
symmetries, while the non-abelian ones present manifest global
symmetries, say $U(1)^{\otimes k}$ or $ SU(2)\otimes U(1)^{\otimes s}$, etc. \cite{elek},
 \cite{vertex}, \cite{local}.
Another important feature of the non-abelian ATFT's is that they always appear in T-dual pairs of IM's,
 related by specific canonical transformations \cite{tdual}.

The established relation between a large class of perturbed CFT's and certain abelian 
and non-abelian ATFT's allows to describe the off-critical properties of the corresponding 2-D
statistical systems in terms of the quantum ATFT's. While the quantization of the abelian ATFT's is
 well developed, it is not the  
case of the non-abelian ATFT's. Their rich symmetry structure  
together with the complicated counterterms needed in their quantum analysis \cite{devega},
make the problem of their quantization difficult and rather different from the  abelian  ATFT's 
\cite{elek}, \cite{fat2}. 
From the other side, due to their
integrability, a  set of stable nonperturbative classical solutions (solitons and breathers) 
of these IM's have been recently
constructed  \cite{pousa}, \cite{elek}, \cite{vertex}, \cite{dorey}.
This suggests that one can follow further the semiclassical methods \cite{DHN}
of quantization of these solutions in order
to derive their nonperturbative particle spectrum, S-matrices, form-factors, etc. \cite{riva} ,\cite{holl-w}.

The aim of the present paper is to study the internal symmetry structure of the soliton solutions of a pair 
of T-dual NA-ATFT's possessing $SL(2)_q\otimes U(1)$ global symmetry, namely:
\br
{\cal L}^{ax} = & -{{k}\o {2\pi }} \(    {1\o {\Delta}}(  {{\bar \pa \psi_1 \pa \chi_1} }(1 + \psi_2 \chi_2 ) 
+ {{\bar \pa \psi_2 \pa \chi_2 }} (1 + \psi_1 \chi_1 + \psi_2 \chi_2 ) \right.   \nonu \\
 & \left.  - {1\o 2}( \psi_2 \chi_1 {{\bar \pa \psi_1 \pa \chi_2 }}+ \chi_2 \psi_1 {{\bar \pa \psi_2 \pa \chi_1 }}))
 -\mu^2 (\psi_1 \chi_1 + \psi_2 \chi_2)\), 
\label{1.1}
\er
where $ \Delta = (1+ \psi_2 \chi_2 )^2 + \psi_1 \chi_1 (1 + {3 \o 4} \psi_2 \chi_2 )  $ and its T-dual counterpart:
\newpage
\br
{\cal L}^{vec} &= & {{k}\o {2\pi }} \( {1\o 2} {{\pa A\bar \pa B + \pa B\bar \pa A}\o {1-AB}} + 
{1\o 2} {{\pa E\bar \pa F + \pa F\bar \pa E}\o {1-EF}} \right. \nonu \\
 & +&\left.  {1\o {2}}{{\pa E\bar \pa A + \pa A\bar \pa E}\o {AE}} + {{\pa E \bar \pa E (1-AB)}\o {E^2(1-EF)}} + \mu^2 (AB+EF -2)\).
\label{1.2}
\er
They represent an $SL(3)$ generalization of the complex SG model \cite{lund} and 
are related to specific integrable perturbation of the 
reduced Gepner  noncompact PF's \cite{gepner}, based on the coset 
\footnote{Since $H_+$ U $H_-$= $GL(2)$ we will offen refer to this coset (for short writing) as "$SL(3)/GL(2)$"}  $H_- \backslash SL(3,R)/H_+ $ 
for $\b = i\b_0, \;\; \b_0^2 = {{2\pi}\o {k}}$, where $H_{\pm}= \{ E_{\pm \a_1}, \l_1 \cdot H, \l_2 \cdot H\}$.

The corresponding  CFT's (i.e.(\ref{1.1}) 
and (\ref{1.2} ) with $\mu=0$)   and their symmetries are studied in Sect.2.
 As  we have shown in our recent paper \cite{vertex},
 these  integrable theories represent 
the simplest example of models  invariant under global transformations from the non-abelian 
q-deformed algebra $SL(2)_q\otimes U(1)$ with $q= \exp (i\b_0^2)$. 
Note that the other known NA-ATFT's as  Homogeneous Sine-Gordon
models \cite{pousa}, $A_n^{(1)}$- dyonic IM's \cite{elek}  and the Fateev's IM's \cite{fat2} admit instead 
only abelian $U(1)^{\otimes s}, \; s=1,2, \cdots $ global
(or local) symmetries. As a consequence of their unusual symmetry structure, both IM's (\ref{1.1}) 
and (\ref{1.2}) possess semiclassical Bohr-Sommerfeld (BS) topological solitons carrying isospin $I$, 
the $U(1)$ charge $Q$ and belonging to certain roots 
of unity representations  
$q^{2(k+3)} = 1$ of $SU(2)_q$ -algebra as  shown in Sect.4.
 The detailed description of 
semiclassical BS - soliton spectrum of both
models presented in Sect.4, allows us to establish the relations between
 their soliton (i.e. strong coupling particles)  spectrum, including masses,
topological and $U(1)$ charges and isospin imposed by the T-duality transformations. We show that the axial and vector 
solitons share the same masses, but their topological charges
are mapped into the $U(1)$ charge $Q$ and the isospin projection $I_3$ and vice-versa. 
This fact  completes the proof of the T-duality of such
IM's.
 
Preliminary discussion concerning the symmetries and the construction of 
soliton solutions of two other integrable models closely related to (\ref{1.1}),
namely the perturbed ungauged $SL(3)$ - WZW model and the off-critical $SL(3)/ {U(1)\otimes U(1)}$
Gepner PF's, is presented in Sect.5. The Appendix contains  the detail
of the derivation of the actions of the  gauged $G/H$ - WZW models studied in Sect.2 and 5.
as well as  of their conserved currents and the explicit form of the constraints, used in the procedure of
Hamiltonian reduction.

\sect{ Gauged $SL(3,R)$ - WZW Models and their massive perturbations}

In this section we derive the actions  of the axial and vector CFT's, based on 
the coset "$SL(3,R)/GL(2,R)$" and study their conformal and internal  $SL(2,R)\otimes U(1)$
symmetries. We also introduce all the ingredients necessary for the construction of their integrable 
perturbations and we  further study a specific massive perturbation of both axial and vector CFT's.

\subsection{Reduced Gepner PF's: axial model }

 The starting point in the derivation and further investigation of the symmetries of the 
IM (\ref{1.1}) is a specific conformal field theory, namely  the axial gauged
 $H_- \backslash SL(3,R)/H_+ $  - WZW model. As explained in  the Appendix  its action can be obtained 
by the standard methods  of refs.\cite{gaw},\cite{or}:
\br
S(g, A, \bar A)^{G/H} = S_{WZW}(g) -{{k}\o {2\pi}}\int dz d\bar z \; Tr ( A\bar \pa g g^{-1}  
 +  \bar Ag^{-1} \pa g +  Ag\bar A g^{-1} + A_0\bar A_0 )
\label{2.1}
\er
where $g\in SL(3,R)$ and $A, \bar A \in H_{\pm}= \{ E_{\pm \a_1}, \l_1 \cdot H, \l_2 \cdot H\}$.
We use a specific Gauss  parametrization (\ref{a0}), (\ref{a3}) and (\ref{a4}) of  the 
matrix fields $A, \bar A$, $g$ and $g_{ax}^f$ in terms of the "physical" fields $\psi_i ,\chi_i $ and $ R_i$ 
 introduced in the Appendix. Following the standard procedure \cite{annals}, and 
integrating over the auxiliary fields 
$A, \bar A$ we derive the effective Lagrangian of the reduced Gepner parafermions in the form (\ref{1.1}) 
with $\mu =0$. It represents further
Hamiltonian reduction of the simplest noncompact Gepner PF's $SL(3,R)/U(1)\otimes U(1)$  by imposing 
the additional constraints, i.e.
\br
J_{\pm \a_1}(z) = \bar J_{\pm \a_1}(\bar z) =0
\label{2.1m}
\er
The chiral  enhanced conformal symmetries of this reduced CFT are generated as usual
 (see refs.\cite{annals},\cite{pol} )
by the remaining $SL(3,R)$ currents, namely 
$J_{\pm \a_2}(z), J_{\pm (\a_1+\a_2)}(z)$ and
$\bar J_{\pm \a_2}(\bar z), \bar J_{\pm (\a_1+\a_2)}(\bar z)$ (see eqn. (\ref{a7})). These currents
can be obtained from the  original $SL(3,R)$ -WZW currents (\ref{a7}) by
imposing the   constraints (\ref{2.1m}), together with two addinional constraints
$J_{\l_1 \cdot H} = J_{\l_2 \cdot H} = 0$, whose explicit form is given in the Appendix. They  indeed 
appear as conserved currents of the gauged 
 $H_- \backslash SL(3,R)/H_+ $ -WZW model. Their specific feature is the presence of 
 nonlocal  terms, due to the elimination of the non-physical fields
 $R_i$ and $\tilde \psi_1$, $\tilde \chi_1$ taking into account the
constraints (\ref{a10}) and (\ref{2.9}), similarly to  the conformal 
NA -Toda models introduced in ref.\cite{annals}.
 The  
classical Poisson brackets (PB) algebra  of these conserved currents can be obtained 
by rewriting  them in terms of the canonical variables,
 $\Pi_{\varphi_a}, \varphi_a = (\psi_i,
\chi_i)$ and next applying the standard canonical PB's:
\br
\{ \varphi_a (x), \Pi_{\varphi_b}(y) \} =- \d_{ab} \d(x-y)
\label{2.1t}
\er
An alternative method \cite{annals}, \cite{pol} consists in imposing the constraints (\ref{2.1m}) and (\ref{a9}) 
directly in the 
$SL(3,R)$-currents algebra transformations or by evaluating the corresponding Dirac brackets.  
The result is the following conformal current algebra of parafermionic type:
\br
\{ J_{\pm \b_i}(\s), J_{\pm \b_j}(\s^{\pr}) \} &=& {{1\o {2k^2}}}\eps (\s -\s^{\pr}) \( J_{\pm \b_i}(\s) J_{\pm \b_j}(\s^{\pr}) +
J_{\pm \b_j}(\s) J_{\pm \b_i}(\s^{\pr})\), \nonu \\
 \{ J_{- \b_i}(\s), J_{\b_j}(\s^{\pr}) \} &=& -{{1\o {2k^2}}}\eps (\s - \s^{\pr}) \( (\b_i \cdot \b_j) J_{-\b_i}(\s) J_{ \b_j}(\s^{\pr}) 
 + \d_{ij} J_{-\b_{s(i)}}(\s)J_{\b_{s(j)}}(\s^{\pr})\) \nonu \\
 &+& \d_{ij} \pa_{\s^{\pr}}\d(\s - \s^{\pr})
 \label{2.2}
 \er
 where $\eps(\s -\s^{\pr})$ is the sign function, $\b_1 = \a_1+\a_2 $ and 
 $\b_2 = \a_2, \; (\b_1^2 =\b_2^2 =2, \b_1 \cdot \b_2 = 1)$
 denote the $SL(3,R)$ roots and $s(1) =2, s(2)=1$.  In order to obtain the complete 
 chiral conformal PF - like algebra of symmetries of the
 gauged WZW-model we have to add to eqns. (\ref{2.2}) the PB's of the stress tensor
 \br
 T= {{1}\o {k}} \(J_{\b_1} J_{-\b_1} + J_{\b_2} J_{-\b_2} \)
 \nonu
 \er
 and its PB's with the currents $J_{\pm \b_i}$ of dimension $\Delta_{cl}(J_{\pm \b_i})=1$:
 \br
 \{T(\s), T(\s^{\pr})\} &=& 2T(\s^{\pr})\pa_{\s^{\pr}}\d(\s - \s^{\pr})+ 
 \pa_{\s^{\pr}}T(\s^{\pr}) \d(\s - \s^{\pr}) - {{k^2}\o {2}}
 \pa_{\s^{\pr}}^3\d(\s - \s^{\pr}), \nonu \\
 \{T(\s), J_{\pm \b_i}(\s^{\pr})\} &=&J_{\pm \b_i}(\s^{\pr})\pa_{\s^{\pr}}\d(\s - \s^{\pr})+ 
 \pa_{\s^{\pr}}J_{\pm \b_i}(\s^{\pr}) \d(\s - \s^{\pr}) 
 \label{2.3}
 \er
 It is worthwhile to note that similar to the PF- and $V_3$- algebras (see eqns. ({32})-({35}) 
 of ref \cite{pl} and Sect. 3 of ref.
 \cite{quantum}) the above PB's algebra can be considered as semiclassical limit ($k \rightarrow \infty $ and
 an appropriate rescaling of the currents) of the 
 reduced Gepner's PF's OPE algebra
 \cite{gepner} of central charge $c= {{4(k-3)}\o {k+3}}$. An important feature of the quantization of the PB 
 algebra (\ref{2.2}) and (\ref{2.1t})
 of the nonlocal currents $J_{\pm \a_1}, J_{\pm(\a_1+\a_2)}$ (\ref{a7}) is that their dimensions 
 $\Delta_{cl}(J)=1$ are  renormalized as follow:
 \br
 \Delta_{1}(J_{\pm \a_2})=1-{1\o k}, \quad 
 \Delta_{2}(J_{\pm (\a_1+ \a_2)})=1+{{k-1}\o k}
 \label{2.5}
 \er
Also the  PB's (\ref{2.2}) and (\ref{2.3}) should  be replaced by the corresponding OPE relations. 
Details concerning this 
enhanced conformal current OPE - algebra and its
 highest weight representations will be presented elsewhere.

The chiral  conformal PF - type algebra  defined by eqns.(\ref{2.2}) and (\ref{2.3}) 
does not exhaust all the symmetries 
of the CFT model (\ref{1.1}) (with $\mu =0$). It is also invariant under the action of an 
additional four parametric set of global
transformations. Its abelian subgroup consists of the following $U(1)\otimes U(1)$ transformations:
\br
\psi_1^{\pr} &=& e^{\b(p_1+p_2)}\psi_1, \quad \chi_1^{\pr} = e^{-\b(p_1+p_2)}\chi_1, \nonu \\
\psi_2^{\pr} &=& e^{\b(-p_1+2p_2)}\psi_2, \quad \chi_2^{\pr} =e^{-\b(-p_1+2p_2)}\chi_2
\label{2.6}
 \er
where $p_i$ are arbitrary constants.
  The corresponding conserved currents ($ \bar \pa I_i = \pa \bar I_i $) have the form:
\br
\b^2 I_1 &=& {{\psi_1 \pa \chi_1}\o {\Delta}}(1+{3\o 2}\psi_2\chi_2)- {{\psi_2 \pa \chi_2}\o {\Delta}}(\Delta_2 + {3\o 2}\psi_1\chi_1), \nonu \\
\b^2 I_2 &=& {{\psi_1 \pa \chi_1}\o {\Delta}} + {{\psi_2 \pa \chi_2}\o {\Delta}}(2\Delta_2 + {3\o 2}\psi_1\chi_1), \nonu \\
\bar I_i &=& I_i (\pa \rightarrow \bar \pa , \; \psi_i \leftrightarrow \chi_i ), \quad \Delta_2 = 1 + \psi_2 \chi_2
\label{2.7}
\er
It is important to note that the constraints equations (\ref{a10}) allows us
 to relate $I_i^{\mu}$ to the topological currents of the ungauged
WZW model
\br
I_i^{\mu} \equiv \eps^{\mu \nu } \b \pa_{\nu}R_i, \quad 2I = I_0 + I_1, \quad 2\bar I = I_0 - I_1
\label{2.8}
\er

The question arises whether the other two constraints $J_{\pm \a_1} = \bar J_{\pm \a_1} = 0$ 
give rise to new conserved currents too. In order to answer it, 
we first rewrite them  in the  following convenient form (see (\ref{a7})):
\br
 \b^2 I_- = {{\psi_2}\o {\Delta}}\(\pa \chi_1 \Delta_2 - 
{1\o 2} \chi_1 \psi_2 \pa \chi_2 \)e^{-{\b \o 2} R_1} \equiv  \pa \tilde \chi, \nonu \\
 \b^2 I_+ ={{\psi_1}\o {\Delta}}\(\pa \chi_2 (1+ \psi_1\chi_1 + \psi_2\chi_2) - 
{1\o 2} \chi_2 \psi_1 \pa \chi_1 \)e^{-{\b \o 2}R_1}\equiv \pa \tilde \psi, \nonu \\
\b^2 \bar I_- = {{\chi_1}\o {\Delta}}\(\bar \pa \psi_2 (1+ \psi_1\chi_1 + \psi_2\chi_2) 
- {1\o 2} \chi_1 \psi_2 \bar \pa \psi_1 \)e^{-{\b \o 2}R_1}\equiv \bar \pa \tilde \chi, \nonu \\
  \b^2 \bar I_+ ={{\chi_2}\o {\Delta}}\(
\bar \pa \psi_1 \Delta_2 - 
{1\o 2} \chi_2 \psi_1 \bar \pa \psi_2 \)e^{-{\b \o 2}R_1}\equiv \bar \pa \tilde \psi, 
\label{2.9}
\er
Their conservation can be easily verified by using the 
equations of motion of (\ref{1.1}) (with $\mu =0$) and eqns. (\ref{2.7}),(\ref{2.8}), i.e. we conclude that
   $\bar \pa I_{\pm} = \pa \bar I_{\pm}$. Hence  the above  nonlocal currents 
$I_{\pm}^{\mu}$ indeed generate new symmetries of the
gauged "$SL(3,R)/GL(2)$" -WZW model. In order to derive the algebra
 of the corresponding conserved charges
\br
Q_l &=& -{i\o { \b}}\int \pa_x R_l dx = \int I_l^0 dx, \nonu \\
Q_+ &=& -{{k}\o {2\sqrt {2} \pi}}\int \pa_x \tilde \psi  dx = - {{k}\o {2 \pi}}\int I_+^0 dx, \nonu \\
Q_- &=& -{{k}\o {2\sqrt {2} \pi}}\int \pa_x \tilde \chi  dx = - {{k}\o {2 \pi}}\int I_-^0 dx, 
\label{2.10}
\er
where $I^0_{\pm} = {{1}\o {\sqrt{2}}}(\bar I_{\pm} + I_{\pm})$, as well as the infinitesimal fields transformations:
\br
\d_{\pm} \psi_i = \{ Q_{\pm} , \psi_i \} \eps_{\pm}
\nonu 
\er
we follow the procedure described above (see also ref.\cite{vertex} ) , 
i.e. we first realize  the currents $I_{\pm , i}^{\mu}$ 
in terms of the canonical 
momenta $\Pi_{\psi_i} = {{\d {\cal L}}\o {\d \pa_0 \psi_i}}$ ( and the
same for $\Pi_{\chi_i}$ ) and next calculate their PB's by using the canonical PB's (\ref{2.1t}):
\br
I_1^0 &=& -{{2\pi }\o {k}} \( \psi_1 \Pi_{\psi_1} - \psi_2 \Pi_{\psi_2} - \chi_1 \Pi_{\chi_1} + \chi_2 \Pi_{\chi_2}\), \nonu \\
I_2^0 &=& -{{2\pi }\o {k}} \( \psi_1 \Pi_{\psi_1} +2 \psi_2 \Pi_{\psi_2} - \chi_1 \Pi_{\chi_1} -2 \chi_2 \Pi_{\chi_2}\), \nonu \\
I_+^0 &=& -{{2\pi }\o {k}} \( \psi_2 \Pi_{\psi_1} - \chi_1 \Pi_{\chi_2}\)e^{-{\b \o 2}R_1}, \nonu \\
I_-^0 &=& -{{2\pi }\o {k}} \( \psi_1 \Pi_{\psi_2} - \chi_2 \Pi_{\chi_1}\)e^{-{\b \o 2}R_1}
\label{2.11}
\er
where 
\br
R_1(x) = -{{2\pi }\o {k}} \int \eps (x-y) \( \psi_1 \Pi_{\psi_1} - \psi_2 \Pi_{\psi_2} - \chi_1 \Pi_{\chi_1} + \chi_2 \Pi_{\chi_2}\)dy
\nonu
\er
The transformations generated by abelian charges $Q_i$ are an infinitesimal form of eqs.(\ref{2.6}):
\br
\d_1 \chi_1 &=& - \chi_1 \eps_1, \quad \d_1 \psi_1 =   \psi_1 \eps_1,\quad \d_1 \psi_2 = -\psi_2 \eps_1,\quad
\d_1 \chi_2 = \chi_2 \eps_1,\nonu \\
\d_2 \chi_1 &=& - \chi_1\eps_2,\quad \d_2 \psi_1 =  \psi_1\eps_2, \quad \d_2 \chi_2 = -2  \chi_2 \eps_2, \quad
\d_2 \psi_2 = 2  \psi_2 \eps_2 
 \label{2.12}
 \er
while the charges $Q_{\pm}$ leads to the following {\it nonlocal} field transformations
 \br
\d_+ \chi_1 &=& {1\o 2} \( \chi_2 e^{-{\b \o 2}R_1} + {1\o 2}\chi_1 \tilde \psi_1 \)\eps_+, \quad \quad
 \d_+ \psi_1 = -{1\o 4}  \psi_1 \tilde \psi_1 \eps_+, \nonu \\
\d_+ \chi_2 &=& -{1\o 4}  \chi_2 \tilde \psi_1\eps_+, \quad \quad
 \d_+ \psi_2 = -{1\o 2} \( \psi_1 e^{-{\b \o 2}R_1} - {1\o 2}\psi_2 \tilde \psi_1 \)\eps_+
 \label{2.13+}
 \er
 and
 \br
 \d_- \chi_1 &=& {1\o 4}  \chi_1 \tilde \chi_1\eps_-, \quad \quad
 \d_- \psi_1 = -{1\o 2} \( \psi_2 e^{-{\b \o 2}R_1} + {1\o 2}\psi_1 \tilde \chi_1 \)\eps_-, \nonu \\
\d_- \chi_2 &=& {1\o 2} \( \chi_1 e^{-{\b \o 2}R_1} - {1\o 2}\chi_2 \tilde \chi_1 \)\eps_-,  \quad \quad
 \d_- \psi_2 = {1\o 4}  \psi_2 \tilde \chi_1\eps_-
 \label{2.13-}
 \er
where $\eps_i = \d \a_i$ and $ \eps_{\pm}$ are arbitrary  
paremeters. One can check by inspection ( and using (\ref{2.9}), (\ref{2.8})) that
transformations (\ref{2.13+}) and (\ref{2.13-}) 
 indeed leave invariant the equations of motion of the gauged WZW model.  

 Finally, the PB algebra of charges $Q_{\pm}, Q_i$ takes the form:
\br
\{ Q_1, Q_{\pm}\} &=& \pm 2 Q_{\pm}, \quad \quad 
 \{ Q_2, Q_{\pm}\} = \mp Q_{\pm}, 
 \label{2.14}
 \er
 and
 \br
 \{ Q_+, Q_{-} \} =- \int_{-\infty}^{+\infty} \pa_x e^{-\b R_1} dx  
 \label{2.15}
 \er
In order to identify the above algebra  with a specific q - deformed algebra, we will 
express the r.h.s. of (\ref{2.15}) in terms of $Q_1$ only. By noting that 
\br
 Q_l = - {{i }\o {\b}}( R_l^+ - R_l^-), \quad \quad    R_l^{\pm} =R_l(\pm \infty ) 
\nonu 
\er
and   next taking into account the following PB's 
\br
\{R_1(x), \pa_y \tilde \chi_1(y)\} = {1\o 2} \eps (x-y) \pa_y \tilde \chi_1 (y), \quad \quad
\{R_1(x), \pa_y \tilde \psi_1(y)\} = -{1\o 2} \eps (x-y) \pa_y \tilde \psi_1 (y).
\nonu
\er
we find that in the limit $x\rightarrow \pm \infty $  the sum $R_1^+ + R_1^-$
 has vanishing PB's with $Q_{\pm}$,i.e.
\br
\{ R_1^+ + R_1^-, \pa_y \tilde \chi_1 (y)\} = 
\{ R_1^+ + R_1^-, \pa_y \tilde \psi_1 (y)\}=0
\nonu 
\er
The above property, together with the  following rescaling  of $Q_{\pm}$ ,
 \br
 E_1= {{e^{{\b \o 4}(R_1^++R_1^-)}\o {(q-q^{-1})}^{1\o 2}}}Q_+, \quad 
 F_1= {{e^{{\b \o 4}(R_1^++R_1^-)}\o {(q-q^{-1})}^{1\o 2}}}Q_-, \quad q=e^{{2i\pi }\o {k}} = e^{i\b^2_0 }
 \label{2.16}
 \er
 allows us to rewrite the PB's (\ref{2.15})  in the well known form \cite{bern} of the $SL(2)_q$ 
 canonical PB's relations,
 \br
 \{E_1, F_1\} = {{q^{I_3} - q^{-I_3}}\o {q-q^{-1}}},
 \label{2.17}
 \er 
 Hence $E_1, F_1$, $Q_1=2I_3$ and $Q = {1\o 3} (Q_1 +2Q_2)$ close into the $SL(2)_q \otimes U(1)$ 
 PB algebra. Our conclusion about the
 symmetries of the gauged WZW-model in consideration is that we have two chiral (left and right) 
 conformal Parafermion  - like algebras
 (\ref{2.2}) and (\ref{2.3}) together with the (nonchiral) $SL(2)_q\otimes U(1)$ algebra 
 (\ref{2.14}) and (\ref{2.17}).  An important
 question concerning the  $SL(2)_q\otimes U(1)$ symmetries of our model is whether exists 
 solutions of the corresponding gauged WZW equations
 of motion that carries nonvanishing  charges $Q_1$ and $Q_2$, otherwise the algebra (\ref{2.17}) becomes  trivial 
 abelian algebra. Since we are interested in the
 integrable perturbations (\ref{1.1}) (with $\mu \neq 0$) that preserves  
 the $SL(2)_q\otimes U(1)$ symmetry, we postpone the answer to this 
 question to Sect.3, where  such $Q_1$ and $Q_2$ charged
 solutions are explicitly constructed.
 
\subsection{T-dual  CFT: vector model}

Following  the procedure described in the Appendix  we derive the Lagrangian of the vector 
gauged $H_- \backslash SL(3,R)/H_+ $ -WZW model in the form (see ref. \cite{tdual} for details): 
\br
 -{{2\pi }\o {k}}{\cal L}_{vec}^{CFT} &=& {1\o 2}\sum_{i=1}^{2} \eta_{ij}\pa \phi_i \bar \pa \phi_j +
 {{\pa \phi_1 \bar \pa \phi_1}\o {t_1^2}}e^{-\phi_1-\phi_2} +\bar \pa \phi_1 \pa ln (t_1) +  
 \pa \phi_1 \bar \pa ln (t_1)  \nonu \\
 &-&
 {\pa \phi_1 \bar \pa \phi_1}{\( {t_2}\o {t_1}\)^2}e^{-2\phi_1 +\phi_2}  
 +  {{\bar \pa (\phi_2 -\phi_1)\pa (\phi_2 - \phi_1)}\o {t_2^2}}e^{\phi_1-2\phi_2}\nonu \\
   &+& \bar \pa (\phi_2 - \phi_1) \pa ln (t_2) +  
 \pa (\phi_2- \phi_1) \bar \pa ln (t_2)  
 \label{svec}
 \er
 where $\eta_{ij} = 2\d_{i,j}-\d_{i,j+1}-\d_{i,j-1}$. As it is shown in ref. \cite{tdual} ( see also refs.\cite{horn}
 and \cite{alvar}
 for other examples of T-dual gauged WZW - models ) the axial CFT
  given by (\ref{1.1}) with $\mu =0$  and
 the vector CFT (\ref{svec}) are T-dual by construction.  The global $U(1)\otimes U(1)$ symmetry 
 (i.e. the presence of two isometric
 directions ) of both models is known to be crucial for their abelian  T-duality. 
  For the axial CFT (\ref{1.1}) they are represented by the
 fields $\Theta_i = ln \({{\psi_i}\o {\chi_i}}\)$, i.e. rewriting eqn. (\ref{1.1}) in the $\s$-model form:
 \br
 \( g^{kl}(\varphi ) \eta^{\mu \nu} + B^{kl}(\varphi )\eps^{\mu \nu }\) \pa_{\mu} \varphi_k \pa_{\nu} \varphi_l
 \nonu
 \er
 the corresponding target space metric $g^{kl}(\varphi ) $ and the antisymmetric tensor 
 $B^{kl}(\varphi )$ are {\it independent} of $\Theta_i$. In order to make evident the 
 isometries of the vector CFT (\ref{svec}) it is
 convenient to change the variables $t_i$ and $\phi_i$ to the following new fields $A,B,E$ and $F$:
 \br
 A= e^{\b (\phi_2 -\phi_1)}, \quad B=A^{-1}\(1- t_2^2e^{\b (2\phi_2-\phi_1)}\), \nonu \\
 E= e^{\b \phi_1}, \quad F=E^{-1}\(1- t_1^2e^{\b (\phi_1+\phi_2)}\),
 \label{2.19}
 \er
 Then
its Lagrangian (\ref{svec}) takes the following simple form:
\br
{\cal L}_{vec}^{CFT} &=& {{k}\o {2\pi}} \( {1\o 2} {{\pa A \bar \pa B +  \pa B \bar \pa A}\o {1-AB}} + 
 {1\o 2} {{\pa E \bar \pa F +  \pa F \bar \pa E}\o {1-EF}} \right. \nonu \\
 &+& \left. {{1}\o {2AE}}(\pa A \bar \pa E + \pa E \bar \pa A) + {{\pa E \bar \pa E (AB -1)}\o {E^2(EF-1)}}\)
 \label{2.20}
 \er
  Now we can choose as isometric coordinates the  following fields
 \br
 \tilde \Theta_2 = -{1\o 2}{\rm ln} A, \quad \quad \tilde \Theta_1 = -{1\o 2}{\rm ln}E,
 \label{2.21}
 \er
 In this context T-duality transformations relating 
 ${\cal L}_{vec}^{CFT}$ and ${\cal L}_{ax}^{CFT}$ can be realized as specific canonical
 (field) transformation $(\Pi_{\tilde \Theta_i}, \tilde \Theta_i) \rightarrow (\Pi_{ \Theta_i},  \Theta_i)$ \cite{alvar},
 \cite{tdual}:
 \br
 \Pi_{\tilde \Theta_i} =-2 \pa_x \Theta_i, \quad \quad \Pi_{ \Theta_i} =-2 \pa_x \tilde \Theta_i
 \label{2.22}
 \er
 while the remaining fields (say, $f_1 = AB, f_2 =EF, a_1 = \psi_1 \chi_1, a_2 = \psi_2 \chi_2 $ ) 
 and their canonical momenta  remain
 unchanged. As a consequence the Hamiltonians of the axial and vector CFT's are identical,i.e.
 ${\cal H}_{vec} = {\cal H}_{ax}$,
 whereas their Lagrangians differ by a total derivative term \cite{tdual},\cite{alvar} :
 \br
 {\cal L}_{vec}^{CFT}(\tilde \Theta_i, f_i) = {\cal L}_{ax}^{CFT}( \Theta_i, a_i) - 
 \tilde \Theta_i (\pa \bar P_i - \bar \pa P_i), 
 \label{2.23}
 \er
where we denote $P_i =\pa \Theta_i, \bar P_i =\bar \pa \Theta_i $ and $\Theta_i = {\rm ln} ({{\psi_i}\o {\chi_i}})$.
 The second term in  (\ref{2.23})
 is nothing but the contribution of the generating function 
 ${\cal F} \sim \sum_{i=1}^2 \eps^{\mu \nu} \pa_{\mu}\Theta_i \pa_{\nu}\tilde
 \Theta_i$ of the canonical transformation (\ref{2.22}).  
 Finally, by applying the Buscher's procedure \cite{buscher} (i.e. taking the Gaussian
 integral  over $P_i, \bar P_i$) one derives the vector CFT ${\cal L}_{vec}^{CFT}$ (\ref{2.20}) starting from 
 the axial one (see ref. \cite{tdual} for details).
 
 Next question to be answered is about the symmetries  of the  $H_- \backslash SL(3,R)/H_+ $ - WZW 
 vector gauged model,  
 given by the Lagrangian (\ref{2.20}). 
  Since by definition T-duality acts as canonical
 transformation (\ref{2.22}) it has to preserve the  corresponding PB's structures\cite{alvar},\cite{tdual}.
  Therefore the CFT 
 (\ref{2.20}) which is T-dual to the axial CFT (\ref{1.1}) ($\mu = 0$) shares  the same PB's algebra of 
 chiral conformal symmetries (\ref{2.2}) and (\ref{2.3}) as well as the global
  $SL(2)_q\otimes U(1)$ given by eqns. (\ref{2.14}) and (\ref{2.15}). One can verify such a
  statement  by direct calculation applying the
  procedure developed in Sect. 2.1 for the derivation of the complete algebra of symmetries 
  of the axial CFT. However there exists an
  important difference in the structure of the global $SL(2)_q\otimes U(1)$  - algebra 
  for the axial and  vector CFT's. Namely, as we have
  shown for the axial model, the abelian subgroup $U(1)\otimes U(1)$ is generated by the 
  charges $Q={1\o 3}(Q_1+2Q_2)$ and $I_3 ={1\o 2}Q_1$ of the
  Noether $U(1)$- currents (\ref{2.7}) giving rise to the  global transformations (\ref{2.6}).  
  Instead, for the vector model we have:
  \br
  \{Q_+, Q_-\} = - \int_{-\infty}^{\infty} dx \pa_x \( {A\o E}\), \quad Q= -{{i}\o {\b^2}}\int \pa_x ln AE, 
  \quad I_3 =-{{i}\o {2\b^2}}\int \pa_x ln {{E}\o {A}}, \nonu \\
  \label{2.24}
  \er
and therefore the abelian subalgebra of the entire algebra of symmetries $SL(2)_q\otimes U(1)$ of the vector model 
is spanned by 
the {\it topological charges} of the isometric fields $\tilde \Theta_1 = - {1\o 2} ln E$ and 
$\tilde \Theta_2 = - {1\o 2} ln A$, defined as usualy
\br
 Q_{vec}^{E} = i\int_{-\infty}^{\infty} \pa_x {\rm ln} \; E \; dx ,\quad\quad\quad
Q_{vec}^{A} = i\int_{-\infty}^{\infty} \pa_x {\rm ln} \; A \; dx 
\nonu
\er
They are indeed different from the Noether charges of the global  
$U(1)\otimes U(1)$ symmetries of ${\cal L}_{vec}$,
\br
A^{\pr} = e^{\b p_1}A, \quad B^{\pr} = e^{-\b p_1}B, \quad 
E^{\pr} = e^{\b p_2}E, \quad F^{\pr} = e^{-\b p_2}F
\label{2.25}
\er
which give rise to the following   nonchiral  $U(1)\otimes U(1)$ currents of the vector model:
\br
J_2^{vec} &=& {1\o 2}\( {{A\pa B - B\pa A}\o {AB-1}} - \pa ln E\), \nonu \\
\bar J_2^{vec} &=& {1\o 2}\( {{B\bar \pa A - A\bar \pa B}\o {AB-1}} + \bar \pa ln E\), \nonu \\
 J_1^{vec} &=& {1\o 2}\( {{E\pa F - F\pa E}\o {EF-1}} - 2{{(AB-1)\pa E}\o {E(EF-1)}} - \pa ln A\), \nonu \\
\bar J_1^{vec} &=& {1\o 2}\( {{F\pa E - E\pa F}\o {EF-1}} + 2{{(AB-1)\bar \pa E}\o {E(EF-1)}} +\bar \pa ln A\)
\label{2.26}
\er 
This phenomena is  a consequence of the main property  of the abelian T-duality transformations 
(\ref{2.22}), namely, it maps the  $U(1)\otimes U(1)$  Noether charges $Q_i^{ax} = \int I_{0,i}^{ax} dx$ of
 the axial model into the topological
charges $Q_{i, vec}^{top} = \int \pa_x \tilde \Theta_i dx$ (those related to the isometric fields only) 
and vice-versa: the topological charges  $Q_{i, ax}^{top} = \int \pa_x  \Theta_i dx$ (of the axial 
isometric fields) to the Noether charges
$Q_i^{vec} = {1\o 2}\int (J_i^{vec} - \bar J_i^{vec})dx$  of the vector model. The first relation is 
encoded in eqns. (\ref{2.7}),
(\ref{2.8}) and (\ref{a10}) and the identification 
\br
-2\tilde \Theta_2 = {1\o 3} (R_2-R_1), \quad -2\tilde \Theta_1 = {1\o 3} (R_2+2R_1)
 \label{2.27}
\er 
 The second relation is based on the following identities:
 \br
 \pa \Theta_1 = 2J_1^{vec}, \quad \pa \Theta_2 = 2J_2^{vec}, \quad \quad
 \bar \pa \Theta_1 = 2\bar J_1^{vec}, \quad \bar \pa \Theta_2 = 2\bar J_2^{vec}
 \label{2.28}
\er 
which are  consequence of eqn. (\ref{a10}) and of the explicit (nonlocal) realization (\ref{3.12}) of the vector model fields $A,B,E,F$ in
terms of the axial model ones.

It should be noted that topological charges can be defined for certain "non-isometric" fields too( see \cite{elek} for relevant
examples), that have properties of "angular" variables. Since T-duality keeps unchanged these fields,  the 
corresponding "non-isometric" topological charges of axial and vector
models do coincide.
 Indeed we can have nontrivial charges only for the fields that have  nonvanishing ( and different ) asymptotics at $ x = {\pm \infty} $.
 As we shall
show in Sect.4.2. below this is the case of the axial model fields $s_1={\rm ln} {{\psi_2 }\o {\psi_1}}$ and 
$s_2={\rm ln} {{ \chi_2 }\o {\chi_1}}$ and
also of their composite 
$ \b r = {\rm ln} \({{\psi_1 \chi_1 }\o {\psi_2 \chi_2}}\) $. The new fields $s_l$ represent the angular part \footnote{observe that the
angular part of the  fields of the  fields $\psi_l$ and  $\chi_l$ belonging to the coset $H_- \backslash SL(3,R)/H_+ $ is related to the Cartan
torus fields $R_i$ as one can see from the eq.(\ref{axang}).}
 $\psi_l$
and  $\chi_l $, i.e. 
\br
\psi_1= e^{{s_1} \o {2}} t_1 ,\quad \psi_2= e^{-{{s_1} \o {2}}} t_1,\quad \chi_1= e^{{s_2} \o {2}} t_2, \quad \chi_2= e^{-{{s_2} \o {2}}}
t_2\nonu\\
s_1={\rm ln}r_1={\rm ln} {{\psi_2 }\o {\psi_1}},\quad s_2={\rm ln}r_2={\rm ln} {{ \chi_2 }\o {\chi_1}},\quad t_1=\sqrt{\psi_1\psi_2},\quad
t_2=\sqrt{\chi_1\chi_2}
\label{angular}
\er

It turns out that
 the one-soliton
solutions of the axial model (\ref{1.1})  give rise to nonzero "non-isometric" topological charges $ Q_{ax,1}^{top}$, $ Q_{ax,2}^{top}$ and
$ Q_{ax}^{top}= Q_{ax,1}^{top}+ Q_{ax,2}^{top}$ ,where we have introduced 
the topological charges related to the fields $s_l$ as
$Q_{ax,l}^{top}=-i\int \pa_x ln (r_l)dx $.
The nonisometric topological charge $Q_{ax}^{top}$ is indeed equal to the
corresponding topological charge of the vector model
\br
Q_{vec}^{top} = i\int \pa_x ln {{EF-1}\o {AB-1}}dx. \nonu
\nonu
\er
The  discussion of their origin, as well as  of the spectrum of these charges for the one-soliton solutions is presented in sect.4 below.

It is important to mention in conclusion that the complete algebra of  symmetries of the vector model  
\br
SL(2)_q\otimes U(1)\otimes \( U(1)\otimes U(1)\)^{Noether}  
\er
contains the $SL(2)_q\otimes U(1)$  algebra having as  Cartan generators the topological charges $Q_{vec}^{E}$ and $Q_{vec}^{A}$ , while 
for the axial model the same algebra contains as Cartan generators 
the $U(1)\otimes U(1)$ Noether charges $I_3 = {1\o 2}Q_1$ and $Q = {1\o 3} (Q_1+2Q_2)$. This is a consequence of the   T-duality of the models that relates the
isometric topological charges of the vector model to the Noether U(1) charges of the axial one.

 \subsection {Perturbed CFT model}
 
The off-critical behaviour of 2-D CFT's is usually described by adding certain 
relevant operators (i.e., combinations
of products of chiral conformal fields of dimension $\Delta <1$) to their CFT's Lagrangians \cite{z1}.  We are 
interested  in massive 
perturbations of the  axial CFT (\ref{1.1}) ($\mu = 0$) (introduced in Sect.2.1),
which  are (i) {\it integrable} and (ii) {\it preserving} the global $SL(2)_q\otimes U(1)$ symmetry. 
 Similar to the standard perturbations
\cite{braz}, \cite{local} of G-WZW model :
\br
{\cal L}^{IM} (g) = {\cal L}^{WZW} (g) +  f_1^{ab} Tr \(gT_ag^{-1}T_b\) - f_2^{ab} Tr \(T_a T_b\)
\nonu
\er
(where $T_a$ are the generators of the Lie algebra $\lie $) we  consider the following perturbation to the Lagrangian
of the gauged $H_- \backslash SL(3,R)/H_+ $  -WZW model
\br
V_{ax} = Tr\( g_{0,ax}^f \eps_- (g_{0,ax}^{f})^{-1} \eps_+\) - Tr \(\eps_+ \eps_-\), \nonu \\
\eps_{\pm} = \mu \l_2 \cdot H^{(\pm 1)} = {{\mu }\o {3}} \( h^{(\pm 1)}_1 + 2h^{(\pm 1)}_2\)
\label{2.30}
\er
where $\l_i$ is the $i^{th}$  fundamental weight of $A_2$ and $h_1$ and $h_2$ denote its Cartan 
subalgebra generators. The  factor group element
$g_{0,ax}^f$ is parametrized as in eqn. (\ref{a4}). The above choice is dictated by the requirement 
\br [\eps_{\pm} , h] = 0, \quad h \in GL(2) = \{E_{\pm \a_1}, h_1, h_2 \}
\nonu 
\er
which guarantees the invariance of the potential 
\br
V_{ax} = \mu^2 \(\psi_1 \chi_1 +\psi_2 \chi_2 \)
\label{2.31}
\er
 under nonlocal field transformations (\ref{2.13+}),(\ref{2.13-}) and the $U(1)\otimes U(1)$ ones given by (\ref{2.6}) as well), i.e.,
\br 
\d _{\pm} V_{ax} = \d_i V_{ax} = 0
\label{2.31m}
\er
Therefore,  the perturbation  (\ref{2.31}) indeed preserves the $SL(2)_q\otimes U(1)$ symmetries of 
the original ${\cal L}_{ax}^{CFT}$
(\ref{1.1}).

  An important question to be answered is about the integrability of the above perturbation.Here
 we shall briefly remind the 
 method for construction of
integrable perturbed CFT's known an Hamiltonian reduction of the two-loop $\hat G$-WZW model \cite{2loop},\cite{elek}.
To this aim  one first consider, instead of the standard finite dimensional $A_2$ - WZW model, its infinite 
dimensional analog \cite{2loop} based on the affine
Kac-Moody algebra $A_2^{(1)}$ and described formally by the same WZW action but 
with group element  $\hat g$ parametrized by  an infinite number of fields.  
  We next define the negative/positive grade subalgebras
$\lie_{<}$ and $\lie_{>}$, according to the approprietly chosen grading operator ${\cal Q}= \hat d$, (i.e. homogeneous gradation)
\br   
[\hat d , E_{\pm \a}^{(l)}] = l E_{\pm \a}^{(l)}, \quad [\hat d , H_i^{(l)}] = l H_i^{(l)},
\label{2.32}
\er
This provides  $A_2^{(1)}$ with a natural  Gauss decomposition 
\br 
\hat G = H_{<} G_0 H_{>}, \quad i.e., \;\; \hat g = Ng_0M
\label{2.33}
\er
where the zero grade subgroup $G_0 = H_< \backslash \hat G/H_> $ is finite dimensional and in the case of $A_2^{(1)}$
is simply $SL(3,R)$ (i.e., $g_0$ is parametrized by eight fields, namely,
 $\tilde \psi_a, \tilde \chi_a, R_i, \;\; a=1,2,3, \;\; i=1,2$ as discussed in
(\ref{a3})). As it well known  the action  of such gauged $G_0 = H_< \backslash \hat G/H_>
 $ - WZW model 
\begin{eqnarray}
S_{G_0}(\hat g,A,\bar{A})&=&S_{WZW}(g_0)
\nonumber
\\
&-&\frac{k}{2\pi}\int d^2x Tr\( A(\bar{\partial}\hat g \hat {g}^{-1}-\epsilon_{+})
+\bar{A}(\hat {g}^{-1}\partial \hat g-\epsilon_{-})+A\hat g\bar{A}\hat {g}^{-1}\) .
\label{2.34}
\end{eqnarray}
where $A\in H_<, \;\; \bar A \in H_>, \;\; g_0\in G_0$  defines  by construction an integrable model characterized 
by the choice of homogeneous gradation ${\cal Q} = d$ and by the
constant elements $\eps_{\pm}$ of grade $\pm 1$(see also \cite{vertex}). In general each 
linear combination of $E_{\pm \b}^{(\pm 1)}$ and  $h_i^{(\pm 1)} ( \; or \;\;
\l_i\cdot H^{(\pm 1)})$ leads to different integrable models, characterized by the subgroup $G_0^0 \subset G_0$ such that $\lie_0^0 = \{ X_0 \in \lie_0 \}$ 
\br 
[\eps_{\pm}, X_0]=0
\nonu
\er
whose structure, in fact, determines  the symmetries of the corresponding IM. The axial IM (\ref{1.1}) 
 is defined by the choice \br
\eps_{\pm} =\mu \l_2 \cdot H^{(\pm 1)} = {{\mu }\o {3}} \( h_1^{(\pm 1)} + 2 h_2^{(\pm 1)}\)
\label{2.35}
\er
and we have $\lie_0^0 = \{E_{\pm \a_1}^{(0)}, h_1^{(0)},  h_2^{(0)} \}$.  Its action 
is obtained by integrating out the
auxiliary fields $A$ and $\bar A$ in the partition function,
\br
Z = \int {\cal D}A {\cal D}\bar A{\cal D} g_0 e^{-S_{G_0}}
\nonu
\er
As a result we obtain
\br
 S_{eff}(g_0) = S_{WZW}^{G_0} (g_0) -{{k}\o {2 \pi}} \int Tr \(\eps_+ g_0\eps_- g_0^{-1}\) d^2x
\label{2.36}
\er
It represents an integrable perturbation of the $G_0=SL(3,R)$ - WZW model ,which  is invariant 
under chiral $G_0^0= SL(2)\otimes U(1)$
transformations. Note that each specific choice of ${\cal Q}$ and $\eps_{\pm}$ leads to IM (\ref{2.36}) 
with different residual symmetries $G_0^0 \otimes  G_0^0$, i.e. $S_{eff}(g_0)$ is invariant under chiral transformations :
\br
g_0^{\pr} = \bar \Omega (\bar z) g_0 \Omega (z)
\nonu
\er
where $\bar \Omega, \Omega \in G_0^0$. Taking into account the corresponding equations of motion ( derived from (\ref{2.36})),
\br
\bar \pa \( g_0^{-1} \pa g_0 \) + [ \eps_-, g_0^{-1} \eps_+ g_0]=0, \quad \quad \pa \(\bar \pa g_0 g_0^{-1} \) - 
[ \eps_+, g_0 \eps_- g_0^{-1} ] =0
\label{2.37}
\er
we conclude that the chiral conserved currents associated to each element $\d_0 \in \lie_0^0$ are given by
\br
J_{\d_0} =  Tr \( \d_0 g_0^{-1} \pa g_0 \) , \quad  \bar J_{\d_0} =  Tr \( \d_0 \bar \pa g_0  g_0^{-1}\), 
\quad \bar \pa J_{\d_0} = \pa \bar J_{\d_0} = 0
\label{2.38}
\er
It is worth to mention that although  $S_{eff}^{G_0}$ in (\ref{2.36}) is invariant under 
chiral $G_0^0 \otimes  G_0^0$ transformations it is not conformal
invariant, due to the second term in its action (\ref{2.36}).

The second step in the construction of the IM representing integrable 
perturbation of the {\it reduced } Gepner PF's  CFT (considered in Sect. 2.1)
consists in further reduction of the IM (\ref{2.36}) by imposing the following set of  additional constraints:
\br
J_{\d_0} = \bar J_{\d_0} = 0, \quad \d_0 \in \lie_0^0
\label{2.39}
\er
This is equivalent of reducing the IM defined on the group $G_0 = SL(3,R)$ to the 
gauged IM defined on the coset $G_0/ G_0^0$ with $\lie_0^0
=  \{E_{\pm \a_1}^{(0)}, h_1^{(0)},  h_2^{(0)} \}$.  The standard procedure  \cite{vertex}, \cite{tdual} 
of realizing this
 gauge fixing of the chiral 
$G_0^0 \otimes  G_0^0$ symmetries is almost identical to the construction of (axial or vector) gauged 
$G/H$ - WZW models summarized in the
Appendix, but  now repeated for the perturbed $G_0$ - WZW model (\ref{2.36}). 

Our next comment is about the symmetries of the IM's (\ref{2.36}), (\ref{1.1}) and (\ref{1.2}). 
 As we have shown (see \cite{vertex} for details)
the perturbed CFT (\ref{2.36}) has chiral $SL(2,R)\otimes U(1)$ symmetries. We have obtained 
the IM's (\ref{1.1}) and (\ref{1.2}) by gauge fixing these chiral symmetries.  The question to 
be answered is about the residual symmetries
of the IM's model (\ref{1.1}) and (\ref{1.2}). We have shown in Sect. 2.1  that the axial CFT 
(i.e., $\mu =0$ limit of IM (\ref{1.1})) is
invariant under global, but nonlocal (in the fields $\psi_i, \chi_i, \; i=1,2$) transformations 
(\ref{2.13+}) and (\ref{2.13-}) as well as under standard global
$U(1)\otimes U(1)$ transformations (\ref{2.12}).  Since our specific choice of the $\eps_{\pm}$ (\ref{2.35}) 
reproduces the perturbation of
this CFT by $V_{ax} = \mu^2 (\psi_1 \chi_1 + \psi_2 \chi_2)$ (or the vector CFT by $V_{vec} = -\mu^2 (AB +EF-2)$)
that have been shown (see eqn. (\ref{2.31m})) to be invariant under (\ref{2.13+}),(\ref{2.13-})  and (\ref{2.12}), then 
the corresponding IM's
(\ref{1.1}) and (\ref{1.2}) representing integrable perturbation of the axial and vector CFT's 
are  indeed invariant under $q$-deformed
$SL(2)_q\otimes U(1)$ algebra.  This can be directly verified by calculating the
 $Q_{\pm}$ and $Q_i$ transformations of the corresponding
Hamiltonians. Finally ,  
using canonical PB's (\ref{2.31}) we verify that
\br
\{ Q_{\pm}, {\cal H}_{ax, vec}^{IM}\} = \{ Q_{i}, {\cal H}_{ax, vec}^{IM}\} =0
\nonu
\er
 and thus confirming once more the  nonlocal symmetries of the IM's (\ref{1.1}) and (\ref{1.2}).

\sect{ Solitons carrying Isospin }
 
 This section is devoted to the construction of the one soliton solutions 
 of the axial and vector IM's (\ref{1.1}) and
 (\ref{1.2}) by means of the vacua Backlund transformation method.

\subsection{Vacua Backlund Transformations}

An important property of the 2-D IM's  ( including NA - ATFT's  in  consideration ) is that  
 one can derive their soliton solutions starting from the constant vacua solution $g_{0 \; vac}$ by applying 
 the Backlund transformation method.  As it is well known \cite{elek} the 1-soliton
  solution $g_{01}$ of eqns. (\ref{2.37}) 
 satisfy the following matrix system of
 $1^{st}$-order differential equations,
 \br
 g_{01}^{-1} \pa g_{01} X - [ g_{01}^{-1} Y g_{0 \; vac}, \eps_-] = 0 \quad \quad 
  \bar \pa g_{01} g_{01}^{-1} Y - [ g_{01} X g_{0 \; vac}^{-1}, \eps_+] = 0
 \label{3.1}
 \er
where the constant matrices $X$ and $Y$  satisfy
\br
[X, \eps_-] = [Y, \eps_+]=0
\nonu
\er
i.e., they have the form:
\br
X = \twomat{{\cal A}}{0}{0}{a_{33}}, \quad \quad Y = \twomat{{\cal B}}{0}{0}{b_{33}}, \nonu \\
{\cal A} = \{a_{ik}, \; i,k= 1,2\} , \quad \quad {\cal B} = \{b_{ik}, \; i,k= 1,2\} \nonu
\er
and the vacua solution of (\ref{2.37}) is given by:
\br
g_{0 \; vac} = e^{\bar a E_{-\a_1}} e^{a_1\l_1\cdot H + a_2\l_2\cdot H} e^{aE_{\a_1}}
\label{gvac}
\er 
with $\bar a$,$ a, a_i, a_{ik}, b_{ik}, a_{33}, b_{33}$ - all arbitrary constants.  
 Since the Lagrangian (\ref{1.1}) and its equations  of motion
(\ref{2.37}) are ${C P}$-invariant,
\br 
\psi_i^{\pr} = \chi_i, \;\; i=1,2, \quad \quad Px = -x, \quad P \pa = \bar \pa 
\label{3.2}
\er
 we require that  the eqns. (\ref{3.1}) are  also invariant under (\ref{3.2}), thus relating the two matrix 
 equations (\ref{3.1}) (and $X$ and $Y$  as well). This provides us with the following simple system of $1^{st}$ order differential equations (DE)
 for the 1-soliton solutions (with $g_{0\; 1}$ parametrized as in (\ref{a3})):
 \br
 \pa \psi_1 &=& -{{m}\o {\g}}\psi_1 \(1+ {{\b^2}\o 2}\psi_2 \chi_2 e^{-{{2\b}\o 3}(R_1 +2R_2)}\)e^{{{\b}\o 3}(R_1 +2R_2)}, \nonu \\
 \pa \chi_1 &=& -{{m}\o {\g}}\chi_1 \(1+ {{\b^2}}\psi_1 \chi_1 +{3\o 2}\b^2 \psi_2 \chi_2\)e^{-{{\b}\o 3}(R_1 +2R_2)}, \nonu \\
 \pa \psi_2 &=& -{{m}\o {\g}}\psi_2 \(1- {{\b^2}\o 2}\psi_1 \chi_1 e^{-{{2\b}\o 3}(R_1 +2R_2)}\)e^{{{\b}\o 3}(R_1 +2R_2)}, \nonu \\
 \pa \chi_2 &=& -{{m}\o {\g}}\chi_2 \(1+ {{\b^2}\o 2}\psi_1 \chi_1 +\b^2 \psi_2 \chi_2\)e^{-{{\b}\o 3}(R_1 +2R_2)}, \nonu \\
 \bar \pa \psi_1 &=& {{m\g}}\psi_1 \(1+ {{\b^2}}\psi_1 \chi_1 +{3\o 2}\b^2 \psi_2 \chi_2\)e^{-{{\b}\o 3}(R_1 +2R_2)}, \nonu \\
 \bar \pa \chi_1 &=& {{m\g}}\chi_1 \(1+ {{\b^2}\o 2}\psi_2 \chi_2 e^{-{{2\b}\o 3}(R_1 +2R_2)}\)e^{{{\b}\o 3}(R_1 +2R_2)}, \nonu \\
 \bar \pa \psi_2 &=& {{m\g}}\psi_2 \(1+ {{\b^2}\o 2}\psi_1 \chi_1 + \b^2 \psi_2 \chi_2\)e^{-{{\b}\o 3}(R_1 +2R_2)}, \nonu \\
 \bar \pa \chi_2 &=& {{m\g}}\chi_2 \(1- {{\b^2}\o 2}\psi_1 \chi_1 e^{-{{2\b}\o 3}(R_1 +2R_2)}\)e^{{{\b}\o 3}(R_1 +2R_2)}, \nonu \\
 \label{3.3}
 \er
where $\g =-\l \({{a_{33}}\o {b_{33}}}\)e^{{1\o 3}(r_1+2r_2)}= e^{b}$.
As consequence of the constraint eqns. (\ref{a10}),(\ref{2.9}) and eqns. (\ref{3.3}) the nonlocal fields $R_i, \tilde \psi , \tilde \chi$
 satisfy the
following $1^{st}$ order DE:
\br
\bar \pa R_1 &=& m\g \b (\psi_1 \chi_1 - \psi_2 \chi_2 )e^{-{{\b}\o 3}(R_1 +2R_2)}, \nonu \\
\bar \pa R_2 &=& m\g \b (\psi_1 \chi_1 +2 \psi_2 \chi_2 )e^{-{{\b}\o 3}(R_1 +2R_2)}, \nonu \\
\pa R_1 &=& - {{m}\o {\g}} \b (\psi_1 \chi_1 - \psi_2 \chi_2 )e^{-{{\b}\o 3}(R_1 +2R_2)}, \nonu \\
\pa R_2 &=& -{{m}\o {\g}} \b (\psi_1 \chi_1 +2 \psi_2 \chi_2 )e^{-{{\b}\o 3}(R_1 +2R_2)}
\label{3.4}
\er 
and 
\br
\pa \tilde \psi = -{{m}\o {\g}} \b \chi_2 \psi_1 e^{-{{\b }\o 6}(5R_1 +4R_2)}, \quad \quad 
\bar \pa \tilde \psi = {{m\g}} \b \psi_2 \chi_1 e^{-{{\b }\o 6}(5R_1 +4R_2)}, \nonu\\
\pa \tilde \chi = -{{m}\o {\g}} \b \psi_2 \chi_1 e^{-{{\b }\o 6}(5R_1 +4R_2)}, \quad \quad 
\bar \pa \tilde \psi = {{m\g}} \b \chi_2 \psi_1 e^{-{{\b }\o 6}(5R_1 +4R_2)}
\label{3.5}
\er
The following algebraic relations also should take place:
\br
\chi_2 \tilde \chi e^{{{\b}\o 2}R_1} - \chi_1 &=& - {{1}\o {\g \rho_1}}\chi_1 e^{{{\b}\o 3}(R_1-R_2)}, \quad \quad 
\psi_2 + \psi_1 \tilde \chi e^{{{\b}\o 2}R_1} =- {{1}\o {\g \kappa_1}}\psi_2 e^{{{\b}\o {3}}(2R_1 + R_2)}
 \nonu \\
\rho_2 \chi_2 &+& \rho_1 \tilde \psi \chi_1 e^{-{{\b}\o 6}(R_1+2R_2)} = {{1}\o {\g}}\chi_2 e^{-{{\b}\o {3}}(2R_1+R_2)}, 
\label{3.52}
\er
 where 
\br
\rho_1 = {{e^{{1\o 3}(2a_1+a_2)}}\o {\l det \; {\cal A}}}\((a_{22}-a_{21}a)(b_{11}+b_{12}\bar a)-a_{21}b_{12}e^{-a_1}\), \nonu \\
\rho_2 = {{e^{{1\o 3}(2a_1+a_2)}}\o {\l det \; {\cal A}}}\((a_{11}a -a_{12})(b_{21}+b_{22}\bar a)+a_{11}b_{22}e^{-a_1}\), \nonu \\
\kappa_1 = {{\l}\o {det \; {\cal B}}}\((b_{12}\bar a+b_{11})(a_{22}-a_{21}a)e^{{{1}\o {3}}(a_1-a_2)}-b_{12}a_{21}e^{-{{1}\o {3}}(2a_1+a_2)}\)
\nonu
\er
As a first step in the construction of solutions of eqns. (\ref{3.3}) and (\ref{3.4}) we derive their "conservation laws": 
\br
\(\g \pa - {1\o {\g}}\bar \pa \) ln {{\psi_i}\o {\chi_i}}= 0, \quad \quad \(\g \pa +{1\o {\g}}\bar \pa \)  ln {(\psi_i \chi_i)}=0
\quad i=1,2, \nonu \\
\(\g \pa +{1\o {\g}}\bar \pa \) (2R_1+R_2) =0, \quad \quad 
\(\g \pa +{1\o {\g}}\bar \pa \) (R_2-R_1) = 0, \nonu \\
\pa  ln \( {{\psi_2}\o {\psi_1}}e^{{{\b}\o 6} (R_1+2R_2)}\) = \bar \pa  ln \( {{\psi_2}\o {\psi_1}}e^{{{\b}\o 6} (R_1+2R_2)} \)=0 \nonu \\
\pa  ln \( {{\chi_2}\o {\chi_1}}e^{{{\b}\o 6} (R_1+2R_2)} \)= \bar \pa  ln \( {{\chi_2}\o {\chi_1}}e^{{{\b}\o 6} (R_1+2R_2)}\) =0
\label{3.6}
\er
The first integrals of the system (\ref{3.3}) and (\ref{3.4}) can be easily obtained from eqns. (\ref{3.3}),(\ref{3.4})
 and (\ref{3.6}):
 \br
 {{\psi_2}\o {\psi_1}}e^{{{\b}\o 6} (R_1+2R_2)} =C_1, \quad \quad {{\chi_2}\o {\chi_1}}e^{{{\b}\o 6} (R_1+2R_2)} = C_2\nonu \\
 2 \sinh ({{\b}\o 3}(R_1+2R_2)) - (\psi_1 \chi_1 + \psi_2\chi_2)e^{-{{\b}\o 3}(R_1+2R_2)} = C_3, \nonu \\
 e^{{{\b}\o 3}(R_2-R_1)}+ C_1C_2 e^{-{{\b}\o 3}(2R_1+R_2)} = C_4
 \label{3.7}
 \er
 where $C_a, a=1,2,3,4$ are arbitrary (complex) constants.  We next observe that all the fields 
 $\psi_i, \chi_i, R_i$ and $\tilde \psi, \tilde \chi$ can be realized in terms of two 
 fields $\phi = {1\o 3}(R_1+2R_2)$ and $u= {{\psi_2}\o {\chi_2}}$ only. For example, we have
 \br
 \chi_1^2 = \({{e^{2\b \phi} -C_3 e^{\b \phi} -1}\o {1+C_1C_2 e^{-\b \phi}}}\){{C_1}\o {C_2}}u^{-1}, \quad \quad 
 \chi_2^2 = \({{e^{\b \phi} - e^{-\b \phi} -C_3 }\o {1+C_1C_2 e^{-\b \phi}}}\){{C_1C_2}}u^{-1}, \nonu \\
 e^{{{\b}\o 2} R_1} = {{1}\o {C_4}}\(e^{{1\o 2}\b \phi} + C_1C_2 e^{-{1\o 2}\b \phi}\), \quad \quad 
 e^{{{\b}\o 2} R_2} = \sqrt{C_4} e^{\b \phi}\(e^{ \phi} + C_1C_2 \)^{-{1\o 2}}
 \label{3.72}
 \er
We next consider the $1^{st}$ order differential equations for the basic fields $\phi$ and $u$ :
\br
\pa_{\rho_+} \(e^{\b \phi}\) &=& m \(1-e^{2\b \phi} +C_3 e^{\b \phi}\), \quad \quad \pa_{\rho_-} \(e^{\b \phi}\) =0, \nonu \\
\pa_{\rho_-}ln (u) &=& - mC_3, \quad \quad \quad \pa_{\rho_+}ln (u) = 0
\label{3.8}
\er
where we have introduced new variables $\rho_{\pm}$,
\br
\rho_+ = x \cosh (b) -t \sinh(b), \quad \quad \pa_{\rho_+} = \sinh (b) \pa_t +\cosh(b) \pa_x, \nonu \\
\rho_- = t \cosh (b) -x \sinh(b), \quad  \pa_{\rho_-} = \cosh (b) \pa_t +\sinh(b) \pa_x, \quad b = ln (\g )
\nonu
\er
Due to the fact that
\br
u=e^{-m C_3 \rho_- +i D}, \quad \quad D= const
\label{3.9}
\er
is time dependent (even in the rest frame, i.e., for $b=0$) an important feature of the finite energy (1-soliton)
 solutions of axial IM
(\ref{1.1}) is that they are not static and one should consider separately  the following two cases
\begin{itemize}
\item periodic in time  particle-like solutions (for $\a \neq 0,\pi $ i.e.$ \sin(\a) \neq 0 $), 
\br
C_3 =-2i \sin (\a), \quad 0\leq \a \leq 2\pi, \nonu \\
\b^2 = -{{2\pi}\o k}, \;\; k>0, \;\; \b=i\b_0, \;\; \b_0 \in R
\nonu
\er
\item nonperiodic in time (unstable) solutions i.e., 
\br
C_3 = -2 \sinh (\a ), \quad \a \in R, \;\; k<0, \;\; \b \in R
\nonu
\er
\end{itemize}
 The second case  is characterized by the fact that the electric charge $Q={1\o 3}( Q_1+2Q_2 )$ and 
 isospin projection $I_3 = {1\o 2}Q_1$  (as well as the
 corresponding topological charges) are
  {\it  not quantized}
 i.e., $Q \in R, I_3 \in R$ and  that the solutions $\psi_i (x,t), \chi_i(x,t)$ are real functions. 
 This choice corresponds to integrable
 perturbation of the  noncompact $SL(3,R)/GL(2,R)$ Gepner PF's (more precisely of $H_- \backslash SL(3,R)/H_+ $-coset model).
 We consider in what follows the first case only,  since similarly to the case of the $A_2$  Abelian Affine Toda IM's (i.e. IM based on
 the principal gradation  \cite {holl-w})  for imaginary coupling $\b = i\b_0$ our  NA Toda IM's(\ref{1.1}) and (\ref{1.2}) 
turns out to admit charged 
topological (and  although complex) soliton solutions with
   real and positive  energy. In the case of the considered NA  affine Toda axial IM (\ref{1.1}) the corresponding  one solitons
 are time - dependent (in the rest frame) and periodic  complex solutions, while for the vector model case (\ref{1.2}) they are
quite similar to the abelian affine Toda case - i.e. they are  static  complex solutions again having real and positive energy. 
They in fact represent solutions of the 
 integrable perturbations of IM's related to the corresponding complex group,i.e. "$SL(3,C)/GL(2,C)$" - WZW
 model. The explicit form of these 1-solitons can be derived from 
 eqns. (\ref{3.3}),(\ref{3.4}),(\ref{3.7}) and (\ref{3.8}):
 \br
 e^{i\b_0 \phi} = e^{-i\a} {{e^f - e^{2i\a}e^{-f}}\o {e^f +e^{-f}}}
 \label{3.10}
 \er
 where $f=m\rho_+ cos (\a) +{1\o 2}X_0$ (with $X_0 =$ const) and 
 \br
 \psi_1 = \pm {{2 cos (\a )}\o {\b_0 (1+C_1C_2e^{i\a})^{{1\o 2}}}}
 {{(e^f -e^{2i\a}e^{-f})^{1\o 2}e^{im\rho_- sin (\a) + {{i}\o 2}D}}\o {(e^f -A_1e^{-f})^{1\o 2}(e^f +e^{-f})}}\({{C_2}\o {C_1}}\)^{{1\o 2}}, \nonu \\
 \psi_2 = \pm {{2 cos (\a )}\o {\b_0 (1+C_1C_2e^{i\a})^{{1\o 2}}}}
 e^{{1\o 2}i\a}{{e^{im\rho_- sin (\a) + {{i}\o 2}D}}\o {(e^f -A_1e^{-f})^{1\o 2}(e^f +e^{-f})^{1\o 2}}}\({{C_2 C_1}}\)^{{1\o 2}}, \nonu \\
 \chi_1 = \pm {{2 cos (\a )}\o {\b_0 (1+C_1C_2e^{i\a})^{{1\o 2}}}}
 {{(e^f -e^{2i\a}e^{-f})^{1\o 2}e^{-im\rho_- sin (\a) - {{i}\o 2}D}}\o {(e^f -A_1e^{-f})^{1\o 2}(e^f +e^{-f})}}\({{C_2}\o {C_1}}\)^{{1\o 2}}, \nonu \\
 \chi_2 = \pm {{2 cos (\a )}\o {\b_0 (1+C_1C_2e^{i\a})^{{1\o 2}}}}
 e^{{1\o 2}i\a}{{e^{-im\rho_- sin (\a) - {{i}\o 2}D}}\o {(e^f -A_1e^{-f})^{1\o 2}(e^f +e^{-f})^{1\o 2}}}\({{C_2 C_1}}\)^{{1\o 2}}, 
 \label{3.11}
 \er
 where $A_1 = e^{2i\a} {{1-C_1C_2 e^{-i\a}}\o {1+C_1C_2 e^{i\a}}}$.  As it is evident from eqn. 
 (\ref{3.11}), $\psi_i , \; \chi_i$ are complex
 periodic functions 
 \br
 \psi_i(x, t+T) = \psi_i(x, t), \quad \quad \chi_i(x, t+T) = \chi_i(x, t)
\nonu
\er
with period $T = {{2\pi}\o {m sin(\a )}}$.  Note that for $\a = \pi l, \;\; l=0,1, \cdots $ 
they become static (i.e. time independent in the
rest frame $b=0$) and complex conjugate ($\psi_i^{*} = \chi_i$).  We postpone the complete study 
of the semiclassical 1-soliton spectrum and
its particle interpretation to Sect.4.

\subsection{Vector Model Solitons}

Topological 1-solitons of the vector model (\ref{1.2}) can be constructed as solutions of the corresponding 
first order equations (\ref{3.1})
written in a parametrization of $g_{01}$ (see eqs. (\ref{a5})-(\ref{a6})) appropriate for 
its description.
There exists 
however a simple way to obtain
these 1-solitons in terms of the axial IM 1-solitons (\ref{3.10}) - (\ref{3.11}).  It is based on the 
following relation between the axial
and vector fields:
\br
 A&=& e^{i{{\b_0}\o 3}(R_2-R_1)}, \quad \quad B= A^{-1}(1+\b_0^2\psi_2\chi_2), \quad \Theta_i = ln \({{\psi_i}\o {\chi_i}}\)\nonu \\
 E&=& e^{i{{\b_0}\o 3}(R_2+2R_1)},\quad \quad \quad F= E^{-1} (1+\b_0^2 \psi_1\chi_1), 
 \label{3.12}
\er
which is obtained  by comparing  the two different parametrizations (\ref{a3}) and (\ref{a5}) 
of the group element $g \in SL(3)$ and  according to the definitions
(\ref{2.19}). It is important to mention that the above nonlocal change  of the field variables (\ref{3.12}) represents 
an alternative form of the abelian T-duality
transformations, similarly to the  axial and vector dyonic IM's considered in ref.\cite{elek}, \cite{dual}.  In order to write the explicit form of the vector IM 1-solitons we need together with 
the axial IM 1-solitons form
(\ref{3.10}) - (\ref{3.11}), the corresponding solutions for the
 nonlocal fields $R_i$ (see eqn. (\ref{3.72})):
\br
e^{i\b_0R_1} &=& {{e^{-i\a}(1+C_1C_2 e^{i\a})^2(e^f -A_1e^{-f})^2}\o {C_4^2(e^f -e^{2i\a}e^{-f})(e^{f}+e^{-f})}},\nonu \\
e^{i\b_0R_2} &=& {{C_4e^{-i\a}(e^f -e^{2i\a}e^{-f})^2}\o {(1+C_1C_2 e^{i\a})(e^{f}+e^{-f})(e^{f} -A_1e^{-f})}}
\label{3.13}
\er
Substituting (\ref{3.13}) and (\ref{3.11}) in eqn. (\ref{3.12}) we derive the 1-solitons of the vector IM (\ref{1.2}):
\br
A&=& {{C_4 (e^f - e^{-f}e^{2i\a})}\o {(1+C_1C_2 e^{i\a})(e^{f} -A_1e^{-f})}}, \quad \quad
E = {{  {(1+C_1C_2 e^{i\a})(e^{f} -A_1e^{-f})}}\o {C_4 (e^{f}+e^{-f})}}e^{-i\a}, \nonu \\
\Theta_1 &=& 2im \rho_- \sin (\a ) +iD + ln {{C_2}\o {C_1}}, \quad \quad \Theta_2 = \Theta_1 - ln \(C_2/C_1\)
\label{3.14}
\er
 Since in the rest frame ($b=0$), $f=mx \cos (\a ) +{1\o 2}X_0$ is time independent, we conclude that
  1-solitons of the vector IM represented
by $A,B,E,F$ according to eqns. (\ref{3.12}) and (\ref{3.14}) are static. As we shall show in Sect.4,
 they carry two nontrivial
topological charges and {\it vanishing } $U(1)\otimes U(1)$ Noether charges,  $Q_i^{vec}=0$.

\sect{Soliton Spectrum and its Symmetries}

The main problem addressed in this section concerns the semiclassical particle - like spectrum
of the one soliton solutions of the axial and vector IM's. 
We derive their masses, topological and Noether charges and  the most important - we identify the specific root of unity
representations of the $SU(2)_q$ algebra  they belong to. 

\subsection {Boundary conditions, vacua and topological charges}
 
 It  is well known, that the topological charges in QFT models are
 closely related to their discrete symmetries  and consequently to the admissible constant
 solutions of the corresponding field equations that provide nontrivial boundary conditions (b.c.'s) for a set of
  fields at $x=\pm \infty$. Their proper definition is as winding number of the map $\phi(x)$ : $R_x$ to $S_1$ or 
more general for matrix valued fields, say $\phi^{ab}(x)$ : $R_x$ to $S_2$, etc. In the simplest case such charges can be 
prescribed to those fields $\phi_i$ that represent angular variables, i.e. for example 
 \br
\phi(+\infty)= \phi(-\infty) + 2\pi n{\cal R} , \quad \quad  n = 0 ,\pm 1,\pm 2,\cdots
\label{comprad}
\er 
where $\cal R$ is the compactification radius of $\phi$ ( the radius of the circle $S_1$ ). It is important to distingush  the case of
arbitrary  $\cal R $  corresponding to winding  $Z$ -type  b.c.'s (as in  sine-Gordon model) from 
the one of "quantized" ${\cal R} = {1 \o k}$ (as in NA - Toda models),
 which gives rise of $Z_k$ -parafermionic (PF) type of b.c.'s. 
 The finite energy solutions such that all the fields in the model have trivial (i.e. zero)
boundary values (and therefore
all the topological charges  vanish), but they carry certain Noether charges, are usualy called "non-topological" solitons \cite{dorey},
\cite{pousa}.
  
  Indeed  the admissible b.c.'s for the fields in a given model should
be derived from its symmetries -  continuous, say $U(1)$ and $SU(2)_q$ or/and discrete - $Z$ or $Z_k$. By
definition  they  represent constant solutions of the models that have zero energy, i.e.,
\br 
 V_{ax} = \psi_1\chi_2 ( {\chi_1\o \chi_2} + {\psi_2\o \psi_1} ) = 0,\quad\quad  V_{vec} = AB + EF - 2 =0
\nonu
\er
 The structure of the {\it classical} vacua manifolds (i.e. zero energy constant solutions) of these models is 
 dictated by their $U(1)\otimes U(1)$
  symmetry. The asymptotics of the one - soliton solutions (\ref{3.11}) and (\ref{3.14})  give us 
  an idea about the type of the  b.c.'s (i.e. vacua) these solitons are connecting.   
For example, for  the boundary conditions
of the fields $A,E,B,F$  of the vector model we obtain  
\br
 A(+\infty)&=&A_1e^{ -2i(\a -{{\pi}\o 2})}A(-\infty) ,\quad B(+\infty)={{1}\o {A_1}} e^{ 2i(\a -{{\pi}\o 2})}B(-\infty) \nonu\\
 E(+\infty)&=&{{1}\o {A_1}}E(-\infty) ,\quad F(+\infty)= A_1F(-\infty),\quad A_1 = e^{2i\a} {{1-C_1C_2 e^{-i\a}}\o {1+C_1C_2 e^{i\a}}}
\label{pfbcs}
\er
 and therefore their vacua ( parametrized by $\a$ and $C_1C_2$ ) represent  a surface $ AB + EF = 2$.
In the case of the axial model, the classical vacua 
are  defined by the nontrivial b.c.'s (at $x=\pm \infty$ ) of the angular fields $s_l$ ,i.e. of $r_1={{\psi_2 }\o {\psi_1}}$ 
and $r_2={ {\chi_2 }\o {\chi_1}}$, according to the definition of the new axial fields (\ref{angular}), 
\br
r _s(+\infty)=\pm C_s e^{-{i\o 2 }\a},\quad r _s(-\infty)=\pm C_s e^{{i\o 2}(\a -{{\pi}\o {2}})},\quad 
r _s(+\infty) = e^{-i(\a -{{\pi}\o {2}})} r _s(-\infty), s=1,2
\label{rbc}
\er
The conditon $V_{ax} = 0$ does not lead to  restrictions on $\a$ and $C_1C_2$, since according to eqs.(\ref{3.11}), 
we always have $\psi_1(\pm \infty) = 0$ and $\chi_2(\pm \infty) =0$. Although we have in both models nontrivial and different b.c.'s 
we can not define proper topological charges without imposing certain
 additional requirements on the values of the parameters $\a$ and $C_1C_2$ (or $A_1$). In fact such (two continuous parameter) family  of
 classical solutions does not represents topological solitons since we can continuously deform each one of them to the vacua (constant, zero
 energy) solution with $\a ={{\pi}\o {2}}$ and $C_1C_2=0$.
As we have already mentioned the topological charges (and topological solitons) require the presence of discrete set of degenerate vacua.
Hence the question to be answered is whether one can further choose the parameters $\a$ and $C_1C_2$  in order
 to realize such   vacua (i.e. b.c.'s)
and the most important question is - which are the symmetries (of the classical or/and quantum theories) that lead to such choice. 
Similar problem \footnote{ it is a generalization of the BS quantization of the SG breather  }
 takes
place in the case of the complex SG and HSG models \cite{dorey},\cite{mirT} and its solution is based on the Bohr-Sommerfeld (BS) 
quantization of the U(1) charge  and the consequencies for the topological charge in the T-dual vector model. More general, various
quantum non-abelian Toda models are known to exibit $Z_k$ symmetries,
obtained by the breaking of the $U(1)$ symmetry in the process of renormalization \cite{fat1},\cite {bard}.
Indeed 
one can choose the parameters in the one - soliton solutions (\ref {3.13}) in order to realize specific PF - type b.c.'s,
\br
 A(+\infty)&=&exp(-2i \pi {l\o k})A(-\infty) ,\quad\quad\quad  B(+\infty)=exp(2i\pi {l\o k})B(-\infty) \nonu\\
 E(+\infty)&=&exp(-2i\pi {{(j_{el}-l)}\o k})E(-\infty) ,\quad \quad  F(+\infty)=exp(2i\pi{{(j_{el}-l)}\o k})F(-\infty)
\label{aebc}
\er
using as an argument the "angular" properties of the Cartan torus fields ($R_i$ or
$\Phi_i$) and  the definitions (\ref{3.12}) of the fields $A,B,E,F$. However  the specific choice of their
compactification radius ${\cal R} = {1 \o k}$ remains a conjecture. The consistent treatement, that makes clear the breaking of the 
$U(1)$ symmetry to $Z_k$ in the procedure of quantization and also the origin of the difference between the classical and quantum vacua,
 has
as its starting point the semiclassical BS -quantization of the time-dependent and periodic 1-solitons (for $sin(\a)\neq 0$) of the axial
model. As we shall see in Sects.4.2 and 4.3. below, the BS procedure selects specific discrete values for $\a$ that lead 
to quantization of their $U(1)$
charge $Q$ (and corresponding topological charges $Q_{ax,l}^{top}$  and $ Q_{ax}^{top}$) 
and determine the compactification radius of a specific combination
${{\b _o}\o {3}} (R_1 +2R_2)$ of the angular fields $R_i$. Such semiclassical BS-solutions interpolate between ceratin (degenerate)
 vacua defined by a {\it
discrete} set of b.c.'s for the fields $s_l$(or $r_l$), they carry rational  topological charges ${ j_{el} \o k}$ 
(related to the angular fields $s_l$) 
and finaly  these semiclassical solutions  can not be deformed to the constant (vacua) solutions. Therefore they represent BS -type of topological solitons
- indeed for very specific (quantised) values of the coupling constant $\b = i\b_0$  and $ \b_0^2 = {{2\pi}\o {k}}$ only 
(where $k=1,2,...$ ).

The BS quantization however  leaves the parameter $C_1C_2$ (and
also $A_1$) arbitrary and therefore it does not reproduce the $Z_k \otimes Z_k$ PF - b.c.'s (\ref{aebc}). Instead it  gives rise  
to certain mixed type $Z_k \otimes U(1)$ of (topological)
b.c.'s. Although in the vector model only the sum of the "topological" charges $Q_{vec}^E $ and $ Q_{vec}^A$ is quantised,  
 the corresponding BS -semiclassical solitons for the vector model are also topological, i.e. they interpolate between a discrete set of b.c.
 and can not be deformed to the vacua solution. In order to obtain topological solitons for the vector
model with two proper topological charges (i.e. both $Q_{vec}^E $ and $ Q_{vec}^A$  quantised) and that interpolate between $Z_k \otimes Z_k$
b.c.'s (\ref{aebc}) one has  to "quantize" the parameter $C_1C_2$ as well. This can be realized by
 requiring either that (1) all the "angular" fields 
( i.e. $R_i$ and related to them $\Phi_i$ ) have equal 
compactificaton radius ${\cal R} = {1\o k}$, or equivalently (2) the isospin $I_3$ to take (half)-integer values (i.e. to consider {\it
unitary} finite dimensional representations  of the larger symmetry group $SU(2)_q$). In the case of the axial
model solitons, as one can see from eq.(\ref{rbc}) 
 the quantization of the $U(1)$ charge $Q$  (i.e. of $\a$) provides the following PF -type
b.c.'s for the "angular" fields $s_l$ (or equivalently of $r_l$ and  $r$), i.e.
\br
r_l(-\infty) = exp({{ i \pi j_{el}}\o k})r_l(+\infty),\quad
{\b_o}r(-\infty) = {\b_o}r(+\infty) + {2\pi j_{el} \o k} 
\label{rbcs}
\er
Therefore such  charged solitons of the axial model  interpolate between discrete degenerate vacua, 
defined by the b.c.'s of the fields $r_l$    at 
$x=\pm \infty$ ( all the other fields  ${\rm ln} ({{\psi_i }\o {\chi_i}})$, or $\psi_1 $ and $\chi_2$ etc. 
have zero asymptotics). The corresponding nontrivial topological charges take discrete values
\br
Q_{ax,l}^{top}=-i\int \pa_x ln (r_l)dx = {{\pi j_{el}}\o {k}}\quad \quad Q_{ax,1}^{top}=Q_{ax,2}^{top} = {{\b_0^2}\o {2}}Q 
\label{noniso}
\er
but the isospin  of these solitons remains continuous. As a consequence the topological 
charges $ Q_{vec}^E $ and $Q_{vec}^A $ of the vector model
are not quantized. Since it is  expected that in the quantum theory the
$U(1) \otimes U(1)$ symmetry to be completely broken to $Z_k \otimes Z_k$ 
\footnote{an indication for this are the discrete $Z_k \otimes Z_k$ symmetries
 of the $\mu =0$ limits of the massive models that are represented  by  quantum CFT's of PF -type }, 
 we shall   impose the condition for  (half)-integer isospin  in this case as well.
  
   It is worthwhile   to mention the particular case of $\sin(\a) = 0$, i.e. $ \a = (j+1)\pi $ (with $j=-1,0$ due to 
  the periodicity of eqs. (\ref{3.11}) in $\a$ ), when the soliton solutions of the axial  
 model (\ref{3.11}) become static and one can no longer use the Bohr-Sommerfeld quantization rule. For these values of
 $\a$  the b.c.'s of the fields $r_l$ (\ref{rbc}) take the simple  form: $r _s(+\infty) = -i(-1)^j r _s(-\infty)$. Such
 solitons carry integer charges 
\br
Q_{ax}^{top} = 2\pi (2j+1)  \quad\quad\quad  Q=(2j+1)k
\nonu
\er
and as  we shall see in sect.4.4. their mass is given by $M ={{2 \mu k}\o \pi}$ (one can also consistenly choose 
their isospin to be (half)-integer).

    The most important property of the finite energy solutions of the considered NA -Toda integrable models is that their  
{\it classical} vacua (defined by the allowed b.c.'s of the classical fields) are described by continuous two parameter surface and the
 corresponding classical solutions (of nontopological type) represent a family  of 
 unstable classical particles. However for specific values of the coupling constant the  corresponding {\it semiclassical} vacua manifold 
 is discrete and one can  
 explicitly  realize semiclassical BS - type topological solitons with quantised (rational) topological charges, which are expected to represent stable 
 (strong coupling) quantum particles. This picture  is quite similar to the original Bohr idea concerning the stability of the atoms - namely the
 electron orbits are quantised, i.e.only a discrete set of  periods (compare with the periods $T = {{2\pi}\o {m sin(\a )}}$ 
 of our solitons that are quantised when $\a$ is
 quantised ) are allowed for the quantum particles, which in fact gives rise to the BS -quantization rules in quantum mechanics
  and in  quantum field theory. It should be noted that these features of the classical and semiclassical solitons of the NA -Toda
  integrable models are quite different from the well known properties of the classical and quantum vacua and (classical and quantum )
solitons in the Sine -Gordon model (as well as of the abelian affine Toda theories) and from the kinks in the broken symmetry $\phi^4$
 model. In all these cases both classical and quantum vacua are discrete(and degenerate) and both classical and quantum solitons are
 topological (for arbitrary values of the coupling constant), i.e.contrary to the considered NA -Toda integrable models,
  the quantization does not change the topological properties of the solitons. Instead in the NA -Toda models the renormalization leads to
  the breaking of the $U(1)$ symmetry   and to quantization of the topological charge, requiring  $Z_k$ -topological b.c.'s
  conditions, that takes place only for specific values of the coupling constant. The difference with the nontopological solitons of the
  complex SG model is that after BS quantization  the solitons of the axial CSG model remains nontopological (there is no topological charge
similar to the one related to the axial fieds $s_l$ ), while in the considered NA -Toda IM's the semiclassical quantization leads
to topological BS - type solitons both in the axial and vector model.

\subsection{ Charges of  axial IM solitons}

According  to its definition (\ref{2.10}) the $U(1)$ charge $Q={1\o 3}(2Q_2+Q_1)$ of 
the 1-soliton solutions (\ref{3.11}) can be realized in
terms of the asymptotic values $R_i(\pm \infty )$ of the nonlocal fields (\ref{3.13}), i.e. we have:
\br
Q = {{2}\o {\b_0^2}}(\a - {{\pi}\o 2}) \equiv -{{i}\o {3\b}}\int \pa_x (2R_2 +R_1)dx
\label{3.15}
\er
Similarly for their isospin projection $I_3$ we obtain,
\br
I_3 = {1\o 2}Q_1 = -{i\o {2\b}} \int dx \pa_x R_1 = 
{{1}\o {\b_0^2}}\( \a - {{\pi}\o 2}+i ln ({{1+C_1C_2e^{i\a}}\o {1-C_1C_2e^{-i\a}}})\)
\label{3.16}
\er
In order to calculate the remaining two charges $Q_{\pm}$ we need the explicit 
form of the nonlocal fields $\tilde \psi$ and  $\tilde \chi$ (given by eqns. (\ref{3.52}) or (\ref{3.5})):
\br
\tilde \psi = \psi_0 +{{C_4}\o {C_1}}\( {{e^{\b \phi}}\o {e^{\b \phi} + C_1 C_2}}\), \quad \quad 
\tilde \chi = \chi_0 +{{C_4}\o {C_2}}\( {{e^{\b \phi}}\o {e^{\b \phi} + C_1 C_2}}\)
\label{3.17}
\er
where $\psi_0, \chi_0$ are certain constant functions of $\rho_1, \rho_2, C_1, C_2$ and $det \; {\cal A}$. 
 Taking into account eqns. (\ref{2.10}) and
(\ref{3.17}) we finally derive 
\br
Q_+ &=& {{\sqrt{2} C_2 C_4 \cos (\a)}\o {\b_0 (C_1 C_2 + e^{-i\a})(C_1 C_2 - e^{i\a})}}, \nonu \\
Q_- &=& {{\sqrt{2} C_1 C_4 \cos (\a)}\o {\b_0 (C_1 C_2 + e^{-i\a})(C_1 C_2 - e^{i\a})}}
\label{3.18}
\er

Next question to be addressed is about the topological charges of the 1-solitons (\ref{3.11}) 
of the axial IM (\ref{1.1}), which are 
related to the "isometric" fields
\br
Q_i^{top, \; ax} = i\int dx \pa_x {\rm ln} ({{\psi_i }\o {\chi_i}}),
\label{3.19}
\er
as well as the ones for the "non-isometric" fields
\br
Q_{ax}^{top} = i\int dx \pa_x {\rm ln} ({{\psi_1 \chi_1 }\o {\psi_2 \chi_2}})= i \b \int dx \pa_x r,\quad 
Q_{ax,l}^{top}=-i\int \pa_x ln (r_l) dx
\label{3.20}
\er
Since we have that by definition  $ Q_{ax}^{top}= Q_{ax,1}^{top}+ Q_{ax,2}^{top}$, only two of 
"nonisometric" topological charges $Q_{ax,l}^{top}$  and $ Q_{ax}^{top}$ are independent.
 As one can easily verify  the "isometric" topological charges  $Q_i^{top, \; ax}$ (\ref{3.19}) of the the 1-solitons
  given by eqns. (\ref{3.11}) 
are zero. An important  feature of the axial
 model (\ref{1.1}) is  that its solitons (\ref{3.11}) have different nonvanishing asymptotics (\ref{rbc}) of the fields $r_l$ and of their
 composite $ \b r = {\rm ln} \({{\psi_1 \chi_1 }\o {\psi_2 \chi_2}}\) $ and therefore they can 
 carry nontrivial "nonisometric" topological charges
 $Q_{ax,l}^{top}$, defined by eq.(\ref{3.20}) above. According to the discussion of Sect.4.1 the classical vacua of the model are
  parametrised by two(continuous) 
 parameters and one have to impose certain "quantization " conditions in order to have discrete degenerate vacua (i.e. b.c.'s of 
 $Z_k$ type for the fields $r_l$) and   properly  defined "nonisometric" topological charges $Q_{ax,l}^{top}$.
  This can be (partialy) achieved  
 by applying   the field theory analog to the Bohr-Sommerfeld rule for quantization of 
classical periodic motions, namely  that  $\int_0^{\tau} p \dot q dt
= 2\pi j_{el}, \;\; j_{el }=\pm 1, \pm 2, \cdots $ to the time -dependent  soliton solutions (see eq.(\ref{3.11}) for
$\sin(\a) \neq 0)$). In this case and for $\b = i\b_0$  these charged 
1-solitons  represent periodic particle-like motion with 
period $\tau = {{2\pi }\o {m sin (\a)}}$.  Therefore we have 
\br
S^{ax} &+& E(v=0) \tau = \int_0^{\tau} dt \int_{-\infty}^{\infty} dx \sum \Pi_{\varphi_a}\dot {\varphi_a} 
= 2\pi j_{el}, \quad \varphi_a = \psi_i, \chi_i, \nonu \\
{\cal H}^{ax} &=& \sum \Pi_{\varphi_a}\dot {\varphi_a} - {\cal L}^{ax}, \quad 
\Pi_{\varphi_a} = {{\d {\cal L}^{ax}}\o {\d \dot {\varphi_a}}}
\label{3.22}
\er
and taking into account $\dot \psi_i, \dot \chi_i$ properties, 
\br
\pa_t \(\psi_i \chi_i\) = 0, \quad \quad \pa_t \( {\rm ln} ({{\psi_i}\o { \chi_i}})\) = 2i m \sin (\a)
\nonu
\er
we derive the following quantization rule:
\br
2\pi j_{el} =-2\pi i\int_{-\infty}^{\infty} dx 
\( \psi_1 \Pi_{\psi_1}- \chi_1 \Pi_{\chi_1} +\psi_2 \Pi_{\psi_2}- \chi_2 \Pi_{\chi_2}\) = 2\pi Q
\label{3.23}
\er
i.e. $Q= j_{el}$.

 An interesting property of the axial model 1-solitons is that the above quantization of the $U(1)$ charge selects
a discrete set of values for the parameter $\a = {\pi \o 2} + {{ {\b_0^2}}\o 2}j_{el}$ . As a consequence we find that the  
b.c.'s for the fields $r_l$ and  $r$,
\br
r_l(-\infty) = exp({{ i \pi j_{el}}\o k})r_l(+\infty),\quad
{\b_o}r(-\infty) = {\b_o}r(+\infty) + {2\pi j_{el} \o k} 
\label{rbcs1}
\er
are of PF - type.
Therefore we have  a discrete set of semiclassical vacua, characterized by the b.c.'s of the fields $r_l$ (\ref{rbcs1}).
The (semiclassicaly quantized) charged solitons  connect two such vacua
  and they carry  topological charges
 $Q_{ax,l}^{top} = {{\pi j_{el}}\o {k}} $ and $Q_{ax}^{top}={{2\pi j_{el}}\o k}$. Observe that  the Noether charge  $Q$ and  the topological
charges $Q_{ax}^{top}$ and $Q_{ax,l}^{top}$  are proportional 
\br
Q_{ax}^{top} = {\b_0^2}Q = 2 Q_{ax,l}^{top}= 2(\a - {\pi \o 2}) = {{2\pi j_{el}}\o k}
\label{3.21}
\er
which is a consequence of  the first two  equations in (\ref{3.7}), i.e. 
\br
i{\b_o}r = {\rm ln} \({{\psi_1 \chi_1 }\o {\psi_2 \chi_2}}\) = i{{\b _o}\o {3}} (R_1 +2R_2) - {\rm ln} C_1 C_2
\nonu
\er
This relation between the Noether charge  $Q$ and  the topological
charge $Q_{ax}^{top}$ for the one - soliton solutions of the axial model 
is not accidental. It follows from the constraints (\ref{a10}) together with  the one - soliton 
equations (\ref{3.3})  and 
(\ref{3.4}) , which relate the fields $r_l$  to an appropriate linear combination of the fields $R_i$. The equivalent relation 
for the vector model connects topological charges  $Q_{vec}^E$ and  $Q_{vec}^A$  with $Q_{vec}^{top}$, i.e. we have
$Q_{vec}^E + Q_{vec}^A = Q_{vec}^{top}$.
Whether similar phenomena takes place for the multi-soliton solutions is an open problem.

It is important to mention that due to the periodicity of the soliton solutions (\ref{3.11}) in $\a$ one can consider the finite interval
$0 \leq \a \leq 2\pi $ only. In fact one can further restrict it to $ -{{\pi}\o {2}} < \a < {{\pi}\o {2}} $, since as one can see from eqs.(\ref{3.11})
and (\ref{3.14}) the asymptotics  of all the fields  depends on the $sign(cos(\a))$. Changing the sign of $cos(\a)$ one exchanges 
the asymptotics at $x = +\infty$ with the ones of $x = -\infty$, say $r_l(+\infty)$  to $r_l(-\infty)$, and therefore it corresponds of
changing the sign of all the charges $Q$ , $Q_{ax,l}^{top}$, etc.,i.e. of solitons with anti-solitons and vice-versa. The values 
$\a=\pm {{\pi} \o {2}}$ (i.e. $j_{el} = 0$) are excluded since they correspond to constant solutions 
(and not to solitons) as one can see from
eqs.(\ref{3.11}) and (\ref{3.14}). This leads us to the  
following restriction on
the range of $j_{el}$ (and also of $Q$ and $Q_{ax,l}^{top}$), namely  $-k< j_{el} <k$ (and
$j_{el} \neq 0$ ), when   solitons and anti-solitons are consider toghether.

Due to  T-duality relations between the
charges in the axial and vector models  the sum of the "vector" topological charges, i.e. 
$Q_{vec}^{top}$ is also quantized. It should be noted  that the  semiclassical BS - quantization 
does not restrict the isospin projection ${j_3}$ to be discrete ,
since the parameter $C_1C_2$ remains arbitrary. As a consequence the vector charges $Q_{vec}^{E}$ and $Q_{vec}^{E}$ 
also have continuum spectrum.
Their quantization, as well as of the isospin, is discussed in the next section.

\subsection{ Charges of  vector IM solitons}

According to the definitions presented in Sect. 2.2, the $U(1)\otimes U(1)$  Noether charges of the vector IM
\br
Q_i^{vec} = \int J_i^{0, \; vec} dx = {1\o 2} \int (J_i^{vec} - \bar J_i^{vec})dx
\nonu 
\er
(see eq. (\ref{2.26}) and (\ref{2.28})) can be realized in terms of the nonlocal fields $\Theta_i$, i.e.,
\br
Q_i^{vec} = \int \pa_x \Theta_i dx
\label{3.24}
\er
Since the fields $\Theta_i$ (see eq(\ref{3.14}) have trivial asymptotic values, the vector model 1-solitons  are chargeless, i.e. 
$Q_i^{vec} = 0$ and  also static (in their rest frame). Due to the relation 
\br 
Q_{vec}^E + Q_{vec}^A = Q_{vec}^{top}
\label{ae}
\er 
between their three topological charges 
\br
Q_{vec}^{top} &=& i\int_{-\infty}^{\infty} \pa_x {\rm ln} \({{EF-1}\o {AB-1}}\)dx =  Q_{ax}^{top}= {\b_0^2}Q, \nonu \\
Q_{vec}^{E} &=& i\int_{-\infty}^{\infty} \pa_x {\rm ln} \; E \; dx = {{i \b}\o {3}}\int_{-\infty}^{\infty} \pa_x (R_2+2R_1)\nonu \\
Q_{vec}^{A} &=& i\int_{-\infty}^{\infty} \pa_x {\rm ln} \; A \; dx = {{i\b}\o {3}}\int_{-\infty}^{\infty} \pa_x (R_2-R_1)
\label{3.25}
\er
we can characterize them by $Q_{vec}^{A}$  and $Q_{vec}^{top}={\b_0^2}Q$  only. Observe that we can realize 
the second topological charge $Q_{vec}^{A}$ in terms of the $U(1)$ Noether 
charges $Q, I_3$  of the axial
model,  
\br
Q_{vec}^{A} = {\b_0^2}({1\o 2}Q - I_3)
\label{3.26}
\er
The charges $Q_{vec}^{top} = Q_{ax}^{top} = {{2\pi j_{el}}\o k}$  ( and also $Q$ ) are already quantized  as we have shown in Sect .4.2.
 The problem to be analysed here is how one can also quantize the isospin $I_3$
or equivalently the topological charge $Q_{vec}^{A}$ , i.e. whether we can choose the free parameter $C_1C_2$ such
that the vector model solitons to interpolate between two different $Z_k \otimes Z_k$ - type b.c.'s. Note that the boundary conditions
for the fields $A,E,B,F$ in general are not of PF-type (see eq.(\ref{aebc})), since we have 
\br
 A(+\infty)&=&A_1e^{ -2i(\a -{{\pi}\o 2})}A(-\infty) ,\quad B(+\infty)={{1}\o {A_1}} e^{ 2i(\a -{{\pi}\o 2})}B(-\infty) \nonu\\
 E(+\infty)&=&{{1}\o {A_1}}E(-\infty) ,\quad\quad  F(+\infty)= A_1F(-\infty)
\nonu
\er
where $A_1 = e^{2i\a} {{1-C_1C_2 e^{-i\a}}\o {1+C_1C_2 e^{i\a}}}$. As we have mentioned in sect.4.1., there are two
equivalent conditions that lead to the quantization of topological charge $Q_{vec}^{A}$ and in the same time  they provide
 b.c.'s of "topological" $Z_k$-type (\ref{aebc}). Namely, one can impose that all  the "angular" fields ($R_i$, $r$ and $\Phi_i$)  
 to have  the same radius of compactification ${\cal R} = {1\o k}$, or equivalently that the isospin takes discrete (half)-integer values.
 Let us consider the consequencies of  the requirement on the isospin $I_3$, i.e. we impose
\br
I_3 = {{k}\o {2\pi}}({1\o 2}Q_{vec}^{top} - Q_{vec}^{A} )= {1\o 2} (j_{el} - 2l) = {1\o 2} j_3, \quad j_3 = 0, \pm 1, \pm 2, \cdots
\label{iso3}
\er
where  $j_3$ is parametrized in terms of $j_{el}$ and an integer $l = 0, \pm 1, \pm 2, \cdots $. As a
result , we find that    the topological charges are given by
 \br
 Q_{vec}^{A} ={{\pi (j_{el}-j_3)} \o k},\quad \quad\quad Q_{vec}^{E} ={{\pi (j_{el}+j_3)} \o k}
 \label{AEtop}
 \er
  The most important consequence  
 of this assumption concerns the form
 of the b.c.'s for the fields $A,E,B,F$. Having in mind the explicit value 
of $I_3$ given by eqn. (\ref{3.16}), i.e. 
\br
j_3 = {{2}\o {\b_0^2}}\( \a - {{\pi}\o 2} +i {\rm ln} ({{1+C_1C_2 e^{i\a}}\o {1-C_1C_2 e^{-i\a}}})\)
\nonu
\er
 and that  $j_{el} = {{2\o {\b_0^2}}}(\a - {{\pi}\o 2})$ we derive the following restriction  :
 \br
C_1C_2 = {{i \sin ({{\b_0^2 l}\o {4}}(j_3-j_{el}))}\o {\sin ({{\b_0^2}\o 4}(j_3+j_{el}))}}
\label{3.27}
\er 
that 
 the parameter $C_1C_2$ should satisfy(for $j_{el} \neq \pm j_3$ and $0<|j_{el}|<k)$. Then the constant $A_1$ 
 that appears in the b.c.'s (\ref{pfbcs}) 
 acquires the following simple form:
\br
A_1 = -e^{i{{\b_0^2}}(j_{el}-l)} = - q^{(j_{el}-l)}= - q^{{1\o 2}(j_{el}+j_3)}, \quad \quad q=e^{{i\b_0^2}}
\label{3.28}
\er
Hence the requirement of  (half)-integer isospin turns out to be equivalent of imposing $Z_k \otimes Z_k$ - PF b.c.'s for the vector model
fields, i.e.
\br
 A(+\infty)&=&exp(-2i \pi {l\o k})A(-\infty) ,\quad\quad\quad  B(+\infty)=exp(2i\pi {l\o k})B(-\infty) \nonu\\
 E(+\infty)&=&exp(-2i\pi {{(j_{el}-l)}\o k})E(-\infty) ,\quad \quad\quad  F(+\infty)=exp(2i\pi{{(j_{el}-l)}\o k})F(-\infty)
\nonu
\er
which are in the origin of the discrete spectrum (\ref{AEtop}) of the topological charges $Q_{vec}^{A}$ and $Q_{vec}^{E}$. In fact,
as we have mentioned in Sect .4.1, the  vacua manifold for the (classical) vector model, 
i.e. those constant solutions(for the fields A,E,B and F) that  also satisfy $AB+EF=2$, 
represents  a  surface parametrized by $\a$ and $C_1C_2$. The conditions of the BS - quantization
of the U(1) charge $Q$ (\ref{3.23}) (in the axial model) together with the isospin quantization (\ref{iso3}) select a finite discrete
set of allowed b.c.'s (\ref{aebc}), i.e.the vacua manifold of the quantum model is discrete and finite dimensional (since $j_{el}$ and $l$
are defined mod k). It is important to mention that the quantization of $\a $  and $C_1C_2$ (and respectively of the parameters $A_1$
(\ref{3.28}) and $ e^{2i\a}$) single out  a discrete set of topological soliton solutions that  interpolate between the different vacua
characterized by the set of $Z_k \otimes Z_k$ - type of b.c.'s (\ref{aebc}). 
  In the case of the axial model the quantization of the $U(1)$ charge $Q$ and of the isospin $I_3$ also selects a discrete set of b.c.'s, 
representing the quantum vacua
and we have a finite number of different soliton solutions(\ref{3.11})(due to the quantization of $\a$ and $A_1$) that connect these vacua.
   
Note that in the corresponding quantum theories for each (quantized) value of the coupling
constant $\b_0^2 ={{2\pi}\o {k}}$ we have different number of "vacua"(and allowed b.c.'s) depending on the value of $k=1,2,3,\cdots$ and as we
shall see in the sect.4.5. below the corresponding soliton solutions belongs to different root of unity representations of the $SU(2)_q$.
  
 The problem of semiclassical 
stability of these soliton solutions
requires further analysis of 
the corresponding "stability" equations (see \cite {holl-w} for the abelian Toda case). We should also note  that 
the promissing method, developed in the
recent paper(\cite{mirT}) for the analysis of the stability  of complex SG solitons, and based on the explicit form of the Bogomol'nyi-like
bounds for the soliton's energy, seems to have a natural extention to the IM's in consideration.
Namely,
 the important ingredient for such analysis - the Backlund type transformations (i.e.the specific first order equations (\ref{3.3})) are
available for the axial and vector model  solitons too. However the problem of classical (and quantum)
 stability of 
the  class of $Z_k \otimes Z_k$ - type 
solitons we have constructed here  is  indeed more  involved and it is out of the scope of the present paper.

\subsection{Soliton Mass  Formula}

In order to permit correct particle interpretation, the 1-soliton solutions (\ref{3.11}) and
(\ref{3.14}) (although complex functions of $x$ for $\b = i \b_0$), should have finite (real and positive)  energy.  
 Similarly to the abelian
 affine Toda \cite{olive}
 and to the $A_n^{(1)}$
dyonic IM's \cite{elek},  the first order soliton equations (\ref{3.3}) and (\ref{3.4}) are 
the main tool in the derivation of the 1-soliton mass
formula.  They allow us  to demonstrate that the stress-tensor components $T^{+}$ and $T^{-}$, 
\br
T^+ &=& {1\o {2\Delta}}\( \pa \psi_2 \pa \chi_2 ( 1+ \b^2 \psi_1 \chi_1 + \b^2 \psi_2 \chi_2) 
+ \pa \psi_1 \pa \chi_1 (1+ \b^2 \psi_2 \chi_2) \right. \nonu \\
&-& \left. 
{\b^2 \o 2}(\psi_2 \chi_1 \pa \psi_1 \pa \chi_2 + \chi_2 \psi_1 \pa \psi_2 \pa \chi_1)\), \nonu \\
T^- &=&T^+ (\pa \rightarrow \bar \pa ), \quad \quad V_{ax} = \mu^2 (\psi_1 \chi_1 +  \psi_2 \chi_2)\nonu \\
T_{00}&=& T^+ + T^- + V, \quad \quad T_{01} = T^+ - T^-
\label{3.29}
\er
can be written as total derivatives, i.e.
\br
T_{00}&=& {{\mu^2}\o 2}  (\psi_1 \chi_1 +  \psi_2 \chi_2)(\g + {1\o \g})^2 = 
{{2 \mu }\o {\b_0^2}} \cosh (b) \pa_x \( e^{i{{\b_0}\o 3}(R_1+2R_2)}\), \nonu \\
T_{01}&=&{{2 \mu }\o {\b_0^2}} \sinh (b)\pa_x \( e^{i{{\b_0}\o 3}(R_1+2R_2)}\)
\label{3.30}
\er
Therefore the mass of the 1-solitons (\ref{3.11}) of the axial model (\ref{1.1}) is given by
\br
M&=& E(b=0) = {{4i \mu }\o {\b_0^2}}\cos (\a ) e^{{i{\b_0 }\o 6}(\phi_+ + \phi_-)}, \nonu \\
\phi_{\pm}  &=& R_1(\pm \infty ) + 2 R_2(\pm \infty )
\nonu
\er 
Taking into account the $R_i(\pm \infty )$ asymptotics derived from eqn. (\ref{3.13}), we find that  
$ie^{{i{\b_0 }\o 6}(\phi_+ + \phi_-)} =1$. Therefore the following formula for the soliton masses takes place :
\br 
M_{j_{el}}^{ax} &=&{{4 \mu }\o {\b_0^2}}|\cos (\a )|\equiv {{4 \mu }\o {\b_0^2}}|\sin ({{\b_0^2 j_{el}}\o 2})|
\label{3.31}
\er
Since $M_{j_{el}}^{ax}$ is independent  of the isospin projection $I_3 = {{j_3}\o 2} = {{j_{el}}\o 2} -l$, 
the solutions labeled by different
values of $l$ form, as we shall see in the next section, multi-component representations of $SU(2)_q$ 
with the same mass $M_{j_{el}}^{ax}$.
  
  The mass of the 1-solitons of the vector model (\ref{1.2}) is given by the same formula (\ref{3.31}) as 
  one  expects, since T-duality
  requires coinciding Hamiltonians, i.e. ${\cal H}_{ax} = {\cal H}_{vec}$. Due to the fact that
   the  electric $U(1)$ charge of the 
  axial model $Q=j_{el}$ is mapped
  into the topological charge of the vector model $Q_{vec}^{top} ={\b_0^2}Q$,  we conclude that
\br 
M^{vec} ={{4 \mu }\o {\b_0^2}}|\sin ({{ Q_{vec}^{top}}\o 2})| = M^{ax}_{j_{el}}
\label{3.32}
\er 
For the specific choice of $ \a = (j+1)\pi $ that leads to static solitons ( both in the axial and the vector models)  
 we find that their mass is given by the simple formula $M ={{2 \mu k}\o \pi}$.

\subsection { Solitons as Representations of $SU(2)_q\otimes U(1)$}

We first note that the 1-soliton mass formula (\ref{3.31}) (in the case $ \a \neq (j+1)\pi $ ) is quite 
similar to the one of the 
sine-Gordon (SG) breathers \cite{DHN} and to the charged
1-solitons of the complex SG model \cite{dorey}.  As one can see from the explicit form 
(\ref{3.11}) ( and (\ref{3.14}))  we have different solutions for
\br 
j_{el} =  \pm 1, \pm 2, \cdots  \pm (k-1) 
\nonu
\er
only, since  the factors $e^{2i \a} = -q^{j_{el}}$  and $A_1 = - q^{(j_{el}-l)}$ are quantized 
in a very specific way, due to periodicity of soliton  solutions in $\a$, as discussed in Sect.4.2. The value $j_{el}= 0$ is excluded
since it corresponds to the constant vacua solutions and not to solitons.  Following  Tseytlin's path integral arguments 
(see Sect. 3 and 4 of ref. \cite{tse}) as in the case of
dyonic IM's \cite{elek} we realize that $\b_0^2$ is renormalized to 
\br
\b_{0, renorm}^2 = {{2\pi }\o {k+3}}, \quad \quad q_{renorm} = e^{i{{2\pi}\o {k+3}}}
\nonu
\er
i.e. $k$ should be replaced by $k+3$. As a consequence, we find that the allowed (renormalized) 
electric charges are given by
\br
j_{el} = \pm 1, \pm 2, \cdots  \pm (k+2)
\label{3.33}
\er 
 In order to derive  the range of the values of $j_3$ (or $I_3$) we observe that  for each fixed value of 
$j_{el}$ as in (\ref{3.33}) the  number of different solutions
 (similar to the species of 1-solitons of the $A_n^{(1)}$ abelian affine Toda
\cite{olive}) is governed by the constant $A_1$  
\br
A_1 (j_{el}, I_3) = -e^{i{{\pi(j_{el}+j_3)}\o {k+3}}}, \nonu
\er 
as one can see from eqns. (\ref{3.11}),(\ref{3.14}) and 
 (\ref{3.28}).  Hence for each fixed value of $j_{el}$ we have $2(k+2)+1$ different values of $A_1$, 
 which provides a set of $2(k+2)+1$ different  soliton solutions
 (with the same mass) characterized by the values of $j_3$ 
 \br
 -(k+2) \leq j_3 \leq (k+2)
\label{3.34}
\er
Therefore the ``renormalized'' solitons, i.e. strong coupling particles of the axial model (\ref{1.1}), form $2k+5= 2I +1$ - component
(i.e., $2(k+3)^{th}$ root of unity) representation of    $SU(2)_q\otimes U(1)$ characterized by its charge $j_{el}$,  its  isospin 
\br
I=k+{5\o 2} = {{2\pi}\o {\b_{0, ren}^2}}- {1\o 2}, \quad i.e., \quad \b_{0, ren}^2 = {{4\pi }\o {2I +1}}
\nonu
\er
and with the  mass, given by
\br
M^{ax} (j_{el},I) = {{m}\o {\pi}}(2I+1) \sin ({{2\pi j_{el}}\o {2I+1}})
\er
Same restrictions for the values of $j_{el}$ and $j_3$ take place for the vector model solitons and again  for each fixed value of 
 $Q_{vec}^{top} = {{2\pi j_{el}}\o {k+3}}$ they form $ 2(k+2)+1$ - multiplets, which components are characterized by the value  of
 topological charge $Q_{vec}^{A} - Q_{vec}^{E} = {{2\pi j_3} \o {k+3}}$, i.e. again  of $j_3$.
 
It should be noted once more that the conclusion concerning the fact that one solitons form certain root of unity representations 
of the $SU(2)_q\otimes U(1)$ algebra is a consequence of the BS - quantization of the $U(1)$ charge and of the requirement to  have
(half)
-integer isospin. As we have shown in sect.4.3. the later is equivalent of imposing $Z_k \otimes Z_k$ - type of 
PF -b.c.'s on the fields of the vector
model, which are expected to take place in the exact quantum theory. In fact we have demonstrated that the parameters in the soliton 
solutions $\a$ and $C_1C_2$
can be consistently chosen in order to have  $Z_k \otimes Z_k$ - type of b.c.'s for the fields and in the same time the one - solitons 
to have finite
energy (mass) and to realize specific finite dimensional representations of the algebra of the symmetries of the models.

\sect{Discussion and Further Developments}
 
 Among the different allowed integrable perturbations of the $SL(3)$ WZW model,
  we have chosen to study the one with the larger
 subgroup of chiral symmetries, i.e. $GL(2)_{left}\otimes  GL(2)_{right}$. 
  By further gauge fixing of  these chiral symmetries, one derives
 \cite{tdual} a pair of T-dual integrable models (\ref{1.1}) and (\ref{1.2}) based 
 on the coset "$SL(3)/GL(2)$", i.e. an integrable perturbation of
 the gauged "$SL(3)/GL(2)$"-WZW model.  They have the remarkable property of being 
 invariant under nonlocal transformations, whose PBs algebra closes
 into the $q$-deformed $SL(2)_q\otimes U(1)$. As a consequence, after semiclassical
 quantization  both theories (for imaginary coupling, i.e. $\b = i\b_0$ ) 
 admit topological solitons (with real and positive energy) carrying isospin and $U(1)$ charge and belonging
 to certain representations of the $SU(2)_q$ as we have shown in Sect. 4.  The explicit
 construction of these solitons allows us to derive
 their semiclassical particle-like spectrum - masses, $U(1)$ charges and isospin.  
 We further
 verify that the one - solitons of the T-dual pair of models share the same masses, but with topological 
 charges mapped into the $U(1)$ charge $Q$ and isospin projection $I_3$.  Thus we have 
 completed the proof of the T-duality of these IM's
 \cite{tdual} on the level of their semiclassical soliton spectrum.  Whether further loop 
 corrections keep or destroy their T-duality 
 remains an an open
 problem, which is closely related to the investigation of their quantum integrability. 
 
 Given the soliton solutions (\ref{3.11}) and (\ref{3.14}) of models (\ref{1.1}) and (\ref{1.2}) , its is
 worthwhile  to mention the crucial role they play in the construction of the solitons of 
 two closely related models, namely 
 \begin{itemize}
 \item Perturbed $SL(3)$  WZW model
 \item Perturbed $SL(3)$ Parafermions, i.e.  $SL(3)/U(1)\otimes U(1)$ model
 \end{itemize}
 It is well known \cite{local},\cite{tdual}  that one can derive 
 the Lagrangian (\ref{1.1}) by first considering an 
 integrable perturbation of the 
 $SL(3)$ - WZW model, such that preserve the invariance under chiral (left 
 and right) $GL(2)$ transformations, namely
\br
{\cal L}_{pert. \; WZW} &=& -{{k}\o {4\pi }}  \( {{1}\o {6}} (2  \pa R_1 \bar \pa R_1 - \pa R_1 \bar \pa R_2 
- \pa R_2 \bar \pa R_1 + 2 \pa R_2 \bar \pa R_2 )   \right. \nonu \\
& + & \left. \pa \tilde \chi_1 \bar \pa \tilde \psi_1 e^{R_1} 
+\pa \tilde \chi_2 \bar \pa \tilde \psi_2 e^{R_2} 
+ (\pa \tilde \chi_3 - \tilde \chi_2 \pa \tilde \chi_1)(\bar \pa \tilde \psi_3 - 
\tilde \psi_2 \bar \pa \tilde \psi_1)e^{R_1 + R_2} -V \)
\label{*}
\er
where $V = \mu^2e^{R_2}\( \tilde \psi_2 \tilde \chi_2 + \tilde \psi_3 \tilde \chi_3\) $. 
As it is explained in  the Appendix (see also Sect.
2 of ref.\cite{vertex})  by gauge fixing the local $GL(2)$  symmetries one  can obtain the Lagrangian (\ref{1.1}).  
 Starting with (\ref{*}) one can gauge
 fix only a part of the local symmetries, namely $U(1)\otimes U(1)$, and the result is the IM (\ref{acaonc})(with $b_1=0$) 
 having rather complicated action. It represents specific integrable
 perturbation of the Gepner's parafermions $SL(3)/U(1)\otimes U(1)$ \cite {pousa}.  The most important consequence of 
 the relationship between IM
 (\ref{1.1}), (\ref{*}) and (\ref{acaonc}) is that one can construct the soliton solutions of the ungauged IM 
 (\ref{*}) or partialy gauged  one (\ref{acaonc})(with  $b_1=0$) 
 by specific conformal dressing  of the solitons of the gauged model (\ref{1.1}) (see \cite{local}).  
  Namely, the fields $\tilde \psi_a, \tilde \chi_a,
 a=1,2,3$ and $R_i, i=1,2$ of (\ref{*}) are related to those $\psi_i, \chi_i$ of (\ref{1.1}) by chiral $U(2)$ 
 gauge transformations
 \br
 \tilde g (\tilde \psi_a, \tilde \chi_a, \tilde R_i) = \bar h(\bar z) g ( \psi_a,  \chi_a,  R_i)h(z), \quad 
 \tilde g, g \in SL(3), \quad h, \bar h \in GL(2)
 \nonu
 \er
 The soliton spectrum of the perturbed $SL(3)$-WZW model (\ref{*}) can be easily derived from the 
 following nonconformal version \cite{local} of
 the standard conformal coset construction \cite{god}:
 \br
 T_{SL(3)}^{IM} = T_{SL(3)/GL(2)}^{IM} + T_{GL(2)}^{CFT}
 \nonu
 \er
 that relates the stress tensors  of the ungauged integrable model (\ref{*}), the gauged one
 (\ref{1.1}) and the $U(2)$-WZW conformal field theory.   
 Similar relations takes place for the $SL(3)/U(1)\otimes U(1)$ (\ref{acaonc}), i.e. independently of the 
 complicated form of its Lagrangian, its solitons (and their semiclasical spectrum) can be easily obtained from
 the ones of IM (\ref{1.1}).

 We belive that our analysis of the symmetry properties, boundary conditions and the spectrum of the solitons of the considered pair of
 T-dual integrable
models represent an usefull input for the further construction of their exact quantum $S$ -matricies. Indeed the most important
open problem is the quantum integrability of such models, which due to their involved renormalization (including complicated
counterterms, etc.) is not straightforward and require the explicit construction of the first few nontrivial higher spin 
(and also fractional
spin (see (\cite {fat1})) quantum conserved charges. 
 
 \vskip .5cm
 
{\bf Acknowledgments.} The authors acknowledge I. Cabrera-Carnero for her assistance at the early stages of this work.
 We are grateful to  CNPq and FAPESP for 
financial support.

\appendix
\section{Appendix }
\subsection{Axial and Vector Gauged WZW Models for $G_0/G_0^0 = SL(3)/GL(2)$}

Consider the specific left-right coset $H_- \backslash SL(3,R)/H_+ $ , where $H_{\pm} \subset SL(3,R)$ are chosen as:
\br
H_{\pm}= \{E_{\pm \a_1}, \; \l_1\cdot H, \; \l_2\cdot H \}, \nonu 
\er
and $\l_i, i=1,2$ are the fundamental weights of $SL(3,R)$.  In order to derive the action 
of the corresponding gauged $H_+  \setminus SL(3,R)/H_-$ - WZW model (see for instance Sect. 2 of \cite{annals},  
and references therein ) we introduce auxiliary
gauge fields \footnote{Notice that our  axial and vector models are obtained from two subsequent gaugings of the WZW model. The first
involves the nilpotent subalgebras \cite{or} generated by $N_{\pm} =\{E_{\pm \a_1} \}$.  The second \cite{gaw} by the Cartan subalgebra of
$SL(2)\otimes U(1)$.  Since the integration over the auxiliary fields are independent of each other, the combination of both gaugings leads
effectively to (\ref{a0}) }
\br
A(z, \bar z) = \sum_{i=1}^{2} a_{0i}(z, \bar z) \l_i\cdot H + a_{-}(z, \bar z) E_{-\a_1} \in H_{-}, \nonu \\
\bar A(z, \bar z) = \sum_{i=1}^{2} \bar a_{0i}(z, \bar z) \l_i\cdot H + \bar a_{+}(z, \bar z) E_{\a_1} \in H_{+}. 
\label{a0}
\er
 Then the action \cite{gaw},  \cite{or}:
\br
S(g, A, \bar A) &=& S_{WZW}(g) -{{k}\o {2\pi}}\int dz d\bar z \; Tr (\eta A\bar \pa g g^{-1}  \nonu \\
 &+&  \bar Ag^{-1} \pa g + \eta Ag\bar A g^{-1} + A_0\bar A_0 )
\label{a1}
\er
is invariant under the following $H_{\pm}$ local gauge transformations
\br
g^{\pr} = h_- g h_+, &  h_- = \g_0(z, \bar z) \g_-(z, \bar z), & h_+ =  \g_+(z, \bar z)\g_0^{\pr}(z, \bar z), \nonu \\
& \g_0^{\pr} = e^{\sum \a_{0i}^{\pr}\l_i\cdot H}, \quad \g_{\pm}= e^{ \a_{\pm }E_{\pm \a_1}}&
\label{a2}
\er
where
\br
A^{\pr} &=& h_- Ah_-^{-1} - \eta \pa h_- h_-^{-1}, \quad A_0^{\pr} = A_0 -\eta \g_0^{-1} \pa \g_0, \nonu \\
\bar A^{\pr} &=& h_+^{-1} \bar A h_+ - h_+^{-1} \bar \pa h_+, \quad \bar A_0^{\pr} = \bar A_0 - \g_0^{-1} \bar \pa \g_0, 
\nonu
\er
where $\eta = 1, \g_0^{\pr} = \g_0$
for axial gauging and  $\eta = -1, \g_0^{\pr} = \g_0^{-1}$ for vector gauging.  
Notice that the auxiliary fields $A$ and $\bar A$ in
(\ref{a0}) act as Lagrange multipliers describing 6 constraints, namely, 
$J_{h_i}= \bar J_{h_i}= J_{\a_1}= \bar J_{-\a_1}=0, \quad i=1,2$.  The Poisson bracket relations,
\br
\lb J_{h_i}(x), J_{\a_1}(y)\rb &=& K_{1,i} J_{\a_1}(x) \d (x-y), \nonu \\
\lb \bar J_{h_i}(x), \bar J_{-\a_1}(y)\rb &=& -K_{1,i} \bar J_{-\a_1}(x) \d (x-y)
\er 
where $K_{ij}$ is the Cartan matrix of $sl(3)$, indicate
 that, together with   the constraints
 $J_{\a_1} = \bar J_{-\a_1}=0$  the gauge fixing conditions (corresponding  to their conjugate momenta),   
$J_{-\a_1} = \bar J_{\a_1}=0$ should also be implemented.  These gauge fixing conditions are consistent with 
\br
\lb J_{\a_1}(x), J_{-\a_1}(y)\rb &=&  J_{h_1}(x) \d (x-y) + k\d^{\pr}(x-y), \nonu \\
\lb \bar J_{\a_1}(x), \bar J_{-\a_1}(y)\rb &=&  \bar J_{h_1}(x) \d (x-y) + k\d^{\pr}(x-y)
\er
Moreover, the constraints $J_{h_i}= \bar J_{h_i}=0, \quad i=1,2$ do not require additional gauge fixing condition since
\br
\lb J_{h_i}(x),J_{h_j} (y)\rb &=&  k\eta_{ij}\d^{\pr} (x-y), \nonu \\
\lb \bar J_{h_i}(x), \bar J_{h_j} (y)\rb &=&   k\eta_{ij}\d^{\pr} (x-y)
\er
i.e. the constraints  appear as gauge fixing conditions to each other.

It is convenient to introduce the following Gauss
parametrization of the group element $g_0$ of $SL(3,R)$:
\br
g_0&=& e^{\tilde \chi_1 E_{-\a_1}} e^{\tilde \chi_2 E_{-\a_2} + \tilde \chi_3 E_{-\a_3}} e^{\phi_1 h_1 + \phi_2h_2 } 
e^{\tilde \psi_2 E_{\a_2} + \tilde \psi_3 E_{\a_3}}e^{\tilde \psi_1 E_{\a_1}} \nonu \\
&=& e^{\tilde \chi_1 E_{-\a_1}}e^{{1\o 2} (R_1 \l_1\cdot H + R_2 \l_2\cdot H)}\( g_{0,ax}^f \) 
e^{{1\o 2} (R_1 \l_1\cdot H + R_2 \l_2\cdot H)}
e^{\tilde \psi_1 E_{\a_1}}
\label{a3}
\er
where $\phi_1 h_1 + \phi_2h_2 = R_1 \l_1\cdot H + R_2 \l_2\cdot H$ and 
\br
g_{0,ax}^f =  e^{\chi_1 E_{-\a_2} +  \chi_2 E_{-\a_1-\a_2}}
e^{\psi_1 E_{\a_2} +  \psi_2 E_{\a_1+\a_2}}, 
\label{a4}
\er
\br
\chi_1(z, \bar z) = \tilde \chi_3 e^{{1\o 2} (R_1+R_2)}, \quad \psi_1(z, \bar z) = \tilde \psi_3 e^{{1\o 2} (R_1+R_2)},\nonu\\ 
\chi_2(z, \bar z) = \tilde \chi_2 e^{{1\o 2} R_2}, \quad \psi_2(z, \bar z) = \tilde \psi_2 e^{{1\o 2} R_2}
\label{axang}
\er
appropriate for the case of axial gauging.  For the vector gauging we start with 
 the same group element,
\br
g_0 = e^{\tilde \chi_1 E_{-\a_1}}e^{{1\o 2} (u_1 \l_1\cdot H + u_2 \l_2\cdot H)}\( g_{0,vec}^f \)
 e^{-{1\o 2} (u_1 \l_1\cdot H + u_2 \l_2\cdot H)} e^{\tilde \psi_1 E_{\a_1}}
\label{a5}
\er
however now the representative of the factor group element is chosen as
\br
g_{0 vec}^f = e^{-t_2E_{-\a_2}} e^{-t_1E_{-\a_1-\a_2} } e^{\phi_1 h_1 + \phi_2h_2} e^{t_2E_{\a_2}} e^{t_1E_{\a_1+\a_2}}
\label{a6}
\er
and $u_1, u_2$ are defined as follows
\br
\tilde \chi_2 e^{-{1\o2}u_2} = -t_2, \quad \tilde \psi_2 e^{{1\o2}u_2} = t_2, \quad 
\tilde \chi_3 e^{-{1\o2}(u_1+u_2)} = -t_1, \quad \tilde \psi_3 e^{{1\o2}(u_1+u_2)} = t_1
\nonu 
\er
The factor group elements $g_{0 ax}^f$ and $g_{0 vec}$ in (\ref{a4}) and (\ref{a6}) are 
constructed to describe how the abelian gaugings are
implemented, axial or vector.
Due to the $H_{\pm}$- invariance we realize that $S(g_0,A, \bar A) = S(g^f_{0 ax, vec},,A^{\pr}, \bar A^{\pr})$.
The corresponding effective actions (\ref{1.1}) for the axial model  and (\ref{1.2}) (with $\mu = 0$)
 for the vector model are obtained by
integrating out  the auxiliary fields $a_{0i}^{\pr}, a_{\pm}^{\pr}$ in the partition function
\br
Z = \int {\cal D}g^f_{0, ax,vec} {\cal D}A^{\pr}{\cal D}\bar A^{\pr}e^{-S(g^f_{0, ax,vec},A^{\pr}, \bar A^{\pr})}
\nonu
\er
(see ref. \cite{tdual} for details).

\subsection{WZW- currents}

In the above parametrization (\ref{a3}) - (\ref{a4}) the $SL(3,R)$-WZW chiral currents ($ J = g_0^{-1}\pa g_0, \;\;
 \bar J = \bar \pa g_0 g_0^{-1},
\;\; \bar \pa J =  \pa \bar J =0$),
\br
J = \sum_{i=1}^{2} J_{\l_i \cdot H}h_i + \sum_{\a}\( J_{\a}E_{-\a}+ J_{-\a}E_{\a} \), \quad \a = \a_1, \a_2, \a_1+\a_2
\nonu 
\er
have the following explicit form:
\br
J_{\a_1+\a_2} &=& (\pa \tilde \chi_3 - \tilde \chi_2 \pa \tilde \chi_1 )e^{R_1+R_2}, \nonu \\
J_{\a_2} &=& (\pa \tilde \chi_2 + \tilde \psi_1 J_{\a_1+\a_2}e^{R_1-2R_2} )e^{-R_1+2R_2},\nonu \\
J_{\a_1} &=& \pa \tilde \chi_1 e^{R_1} - \tilde \psi_2 J_{\a_1+\a_2},\nonu \\
J_{-\a_1} &=& \pa \tilde \psi_1 - \tilde \psi_1^2 \pa \tilde \chi_1 e^{R_1} + \pa \tilde \chi_2 (\tilde \psi_1 \tilde \psi_2 - \tilde
\psi_3)e^{R_2} \nonu \\
& & + J_{\a_1+\a_2}(\tilde \psi_1 \tilde \psi_2 - \tilde \psi_3) \tilde \psi_1 + \tilde \psi_1 \pa R_1,\nonu \\
J_{-\a_2} &=&-\tilde \psi_2^2 \pa \tilde \chi_2 e^{-R_1+2R_2} - \tilde \psi_2\pa R_1 + \pa \tilde \psi_2 + 2 \tilde \psi_2 \pa R_2, \nonu \\
J_{-\a_1-\a_2} &=& \pa \tilde \psi_3 +\tilde \psi_3 (\pa R_1 + \pa R_2)- \tilde \psi_3^2 J_{\a_1+\a_2} - \tilde \psi_2 \tilde \psi_3 \pa
\tilde \chi_2 e^{-R_1+2R_2} -\tilde \psi_1 J_{-\a_2}, \nonu \\
J_{\l_1 \cdot H} &=& {1\o 3} (2\pa R_1 + \pa R_2) - \tilde \psi_1 \pa \tilde \chi_1 e^{R_1} + (\tilde \psi_1 \tilde \psi_2 - \tilde \psi_3)
J_{\a_1+\a_2},\nonu \\
J_{\l_2 \cdot H} &=& {1\o 3} (\pa R_1 + 2\pa R_2) - \tilde \psi_2 \pa \tilde \chi_2 e^{R_2} - 
\tilde \psi_3 J_{\a_1+\a_2}
\label{a7}
\er
and $\bar J = J(\pa \rightarrow \bar \pa, \tilde \psi_j \rightarrow \tilde \chi_j,\;  j=1,2,3)$. 
 Note that both the axial and the vector gauged WZW models corresponds to imposing the following constraints on the $SL(3,R)$ currents:
 \br
 J_{\pm \a_1} &=& \bar J_{\pm \a_1} = 0, 
 \label{a8}\\
 J_{\l_i\cdot H} &=& \bar J_{\l_i\cdot H} = 0,\;\;  i=1,2
 \label{a9}
 \er
The explicit form of constraints (\ref{a9}) in the parametrization (\ref{a3}) and  (\ref{a4}) is as follows (see \cite{vertex}):
\br
 \pa R_1 &=& {{\psi_1 \pa \chi_1} \o {\Delta}}( 1+{3\o 2}\psi_2 \chi_2)  - {{\psi_2 \pa \chi_2 }\o {\Delta}}( \Delta_2 + 
 {{3\o 2}} \psi_1 \chi_1), \nonu \\
 \pa R_2 &=& {{\psi_1 \pa \chi_1  }\o {\Delta}} + {{\psi_2 \pa \chi_2 }\o {\Delta}}( 2 \Delta_2 + {{3\o 2}} \psi_1 \chi_1), \nonu \\
\bar  \pa R_1 &=&  
{{\chi_1 \bar \pa \psi_1 } \o {\Delta}} ( 1+{3\o 2}\psi_2 \chi_2)  - {{\chi_2 \bar \pa \psi_2 }\o {\Delta}}( \Delta_2 + 
{{3\o 2}} \psi_1 \chi_1),\nonu \\
 \bar \pa R_2 &=&
{{\chi_1 \bar \pa \psi_1  }\o {\Delta}} + {{\chi_2 \bar \pa \psi_2 }\o {\Delta}}( 2 \Delta_2 + {{3\o 2}} \psi_1 \chi_1) 
 \label{a10}
\er 
where
$ \Delta = (1+ \psi_2 \chi_2 )^2 + \psi_1 \chi_1 (1+ {3\o 4} \psi_2 \chi_2 ), \quad \Delta_2 = 1+ \psi_2 \chi_2$.

Eqns. (\ref{a10}) can be considered as definition of the nonlocal fields $R_i$ of the axial model. 
 The constraints (\ref{a8}) are given by
eqn. (\ref{2.9}) with $\tilde \chi_1 = \tilde \chi, \;\;  \tilde \psi_1 = \tilde \psi$.

\subsection{Axial Gauged WZW Model for $G_0/G_0^0 = SL(3)/U(1)\otimes U(1)$}
 
In order to  derive the axial gauged action for the  $SL(3)/U(1)\otimes U(1)$ - WZW model we shall 
employ the same arguments as in the case of the $G_0/G_0^0 = SL(3)/GL(2)$ -  coset model discussed above, but now
with  
\br
H_+ =\{ \l_1\cdot H, \l_2\cdot H\}
\nonu
\er
and consider the gauged WZW action given by (\ref{a1}) with $\eta = +1$.  Following the arguments of refs. \cite{elek}
\cite{annals} we define 
\br
A= A_0 = a_{01}\l_1\cdot H + a_{02}\l_2\cdot H, \quad \quad 
\bar A = \bar A_0= \bar a_{01}\l_1\cdot H + \bar a_{02}\l_2\cdot H
\label{b11}
\er
and 
\br
 g_{0,ax}^f =  e^{\chi_3 E_{-\a_1} +  \chi_2 E_{-\a_2} +\chi_1 E_{-\a_1-\a_2}}
e^{\psi_3 E_{\a_1} +  \psi_2 E_{\a_2}+ \psi_1 E_{\a_1+\a_2}}, 
\label{b4}
\er
Since the action (\ref{a1}) is invariant under transformations (\ref{a2}),  
 we find $S(g_0,A, \bar A) = S(g^f_{0 ax}, A^{\pr}, \bar A^{\pr})$.
 By direct calculation we find
 \br
 Tr ( A_0\bar \pa g_0^f (g_0^{f})^{-1}  
 + \bar A_0(g_0^{f })^{-1}\pa g_0^f +  A_0g_0^f\bar A_0 (g_0^{f })^{-1} + A_0\bar A_0 )
 =\bar {\it a} M {\it a} +\bar N {\it a} + \bar {\it a} N
\label{b5}
\er
 where
 \br
M= \twomat{{4\o3} +\psi_3\chi_3 + (\psi_1 -{1\o 2}\psi_2\psi_3)(\chi_1 - 
 {1\o 2}\chi_2\chi_3)}{-{2\o 3} -\psi_3\chi_3 + \chi_2\chi_3(\psi_1
 -{1\o 2}\psi_2\psi_3)}{-{2\o 3} -\psi_3\chi_3+ \psi_2\psi_3(\chi_1-
 {1\o 2}\chi_2\chi_3)}{{4\o3} +\psi_3\chi_3 +\psi_2\chi_2+\psi_1\chi_1}, \nonu \\
 {\it a} = \twocol{a_{01}}{a_{02}}, \quad  N = \twocol{ N_1}{ N_2}, \quad 
 \bar {\it a} = \twovec{\bar a_{01}}{\bar a_{02}}, \quad \bar N = \twovec{\bar N_1}{\bar N_2}\quad 
 \er
 \br
 \bar N_1 &=& -\bar \pa \psi_1 (\chi_1 -{1\o 2}\chi_2\chi_3)-{1\o 2}\psi_3 \bar \pa \psi_2
 (\chi_1 - {1\o 2}\chi_2 \chi_3)-{1\o 2}  \bar \pa \psi_3(\chi_3 -{1\o 2}\psi_2(\chi_1 -{1\o 2} \chi_2\chi_3)), \nonu \\
 \bar N_2 &=&-\bar \pa \psi_1 \chi_2\chi_3 -\bar \pa
 \psi_2\chi_2 (1+ {1\o 2} \chi_3\psi_3)+ \bar \pa \psi_3 \chi_3 (1+{1\o 2}\psi_2\chi_2), \nonu \\
 N_1 &=& -\pa \chi_1(\psi_1 -{1\o 2}\psi_2\psi_3)-{1\o 2}\chi_3\pa \chi_2 (\psi_1 - {1\o 2}\psi_2 \psi_3) 
 -\pa \chi_3 (\psi_3 -{1\o 2}\chi_2(\psi_1
 -{1\o 2} \psi_2\psi_3)) , \nonu \\
 N_2 &=&-\psi_3\psi_2 \pa \chi_1 -\psi_2 \pa \chi_2 (1+{1\o 2} \psi_3 \chi_3)+ \psi_3 \pa \chi_3 (1+{1\o 2} \psi_2 \chi_2)
 \er
 Integrating over the matrix valued auxiliary fields $\bar {\it a}, {\it a}$ we arrive at the following effective
 Lagrangian :
 \br
{\cal L} = {{\pa \chi_i \bar \pa \psi_j}\o {24 \Delta}}\Delta_{ij}
\label{acao}
\er
where 
\br
\Delta_{11} &=& 32(1 + \chi_3 \psi_3 + \chi_2 \psi_2  + {3\o 4} \chi_2 \psi_2 \chi_3 \psi_3), \nonu \\
\Delta_{12} &=&16 \psi_3 (1 + \chi_3 \psi_3) - 8\chi_2 (2 \psi_1 + 3 \chi_3 \psi_1 \psi_3 + \psi_2 \psi_3) \nonu \\
\Delta_{13} &=& -16 \psi_2 (1 + \chi_2 \psi_2) + 6 \chi_3^2  \psi_3 (2 + \psi_2) (2 \psi_1 + \psi_2 \psi_3) \nonu \\
&+&
4\chi_3  \psi_1 (4 + 4 \psi_2 + 3 \chi_2 \psi_2^2 ) 
+ 2\psi_2 \chi_3\psi_3\( 4 - 3 \chi_2 \psi_2 (2 + \psi_2)\) , \nonu \\
\Delta_{21} &=&  -16 \chi_1 \psi_2 +16 \chi_3^2  \psi_3 -8 \chi_3 (-2 + \chi_2 \psi_2 + 3 \chi_1 \psi_2 \psi_3), \nonu \\
\Delta_{23} &=& 8 \chi_1 \psi_2^2  + 4 \chi_3 \psi_2 \(6 + 6 \chi_1 \psi_1 + \psi_2(2 + \chi_2) \) \nonu \\
 &+&
3  \chi_3^3  \psi_3(2 + \psi_2)  (2 \psi_1 + \psi_2 \psi_3) 
   +2 \psi_1 \chi_3^2\( 4 + (4 - 6 \chi_2) \psi_2 - 3 \chi_2 \psi_2^2\)\nonu \\
 &+&
\psi_2 \psi_3\chi_3^2\( 28 + 6 (2 + \chi_2) \psi_2 + 3 \chi_2 \psi_2^2 \), \nonu \\
\Delta_{22} &=& 32 - 8\chi_2 \chi_3 \psi_1 + 40 \chi_3 \psi_3 + 8\chi_2 \chi_3 \psi_2 \psi_3 + 8\chi_3^2  \psi_3^2  
  -8\chi_1 \psi_2 \psi_3 + 8\chi_1\psi_1 (4 + 3 \chi_3 \psi_3), \nonu \\
  \Delta_{32} &=&  -8 \chi_1 \psi_3^2  + 4\chi_2 \psi_3 (2 + 6 \chi_1 \psi_1 + \chi_3 \psi_3) +
 4\chi_2^2  (2 \psi_1 + \psi_2 \psi_3), \nonu \\
 \Delta_{31} &=& -16 \chi_2^2 \psi_2 -16 \chi_1 \psi_3 -4\chi_2 (4 - 2 \chi_3 \psi_3 + 6 \chi_1 \psi_2 \psi_3), \nonu \\
 \Delta_{33} &=& 8 \chi_1 \psi_1 (4 + 3 \chi_2 \psi_2) + 8\chi_1\psi_2 \psi_3 + 32 + 8\chi_3 (2 + \psi_2) \psi_3 \nonu \\
 & +&
 8 \chi_2^2 \psi_2^2  + 3 \chi_2^2 \psi_2\chi_3 (2 + \psi_2) (-2 \psi_1 + \psi_2 \psi_3) \nonu \\
 &+&
   40 \chi_2 \psi_2 + 3 \chi_2\chi_3^2  \psi_3(2 + \psi_2)  (2 \psi_1 + \psi_2 \psi_3)\nonu \\
  & &-8  \chi_3 \chi_2 \psi_1 (1 + \psi_2) +4\chi_3 \chi_2 \psi_3\psi_2(8 + 3 \psi_2) 
 \er
 and 
 \br
 \Delta &=& {4\o 3} +{4\o 3}(\psi_1\chi_1 + \psi_2\chi_2 + \psi_3 \chi_3) + {1\o 2}(\psi_1 \chi_2 \chi_3 + \chi_1 \psi_2 \psi_3 )(\psi_3
 \chi_3 - \psi_2 \chi_2)\nonu \\
 & +& {1\o 4}\psi_2 \chi_2 \psi_3 \chi_3 (\psi_3\chi_3 + \psi_2 \chi_2)
 \er
 
 A potential invariant under the $U(1)\otimes U(1)$ transformations generated by $H_+$ may be 
 written in terms of the grade $\pm 1$ constant
 operators $\eps_{\pm}$ as follows :
\br
 V = Tr \( \eps_+g_0^f \eps_- (g_0^{f})^{-1}\) &=& (b_1^2\l_1^2 + 2b_1b_2 \l_1\cdot \l_2 +b_2^2\l_2^2) + b_1^2 \psi_2 \chi_2 +
 + b_2^2 \psi_3 \chi_3 \nonu \\
 &+&(b_1+b_2)^2 (\chi_1 - {1\o 2} \chi_2\chi_3) (\psi_1 - {1\o 2} \psi_2\psi_3)
 \er
 where 
 \br
 \eps_{\pm} = b_1 \l_1 \cdot H^{(\pm )} + b_2 \l_2 \cdot H^{(\pm )} 
\er
 such that the non - conformal model  can be described by
 \br
{\cal L} = {{\pa \chi_i \bar \pa \psi_j}\o {24 \Delta}}\Delta_{ij} -V
\label{acaonc}
\er
It represents an integrable perturbation of the  non-compact Gepner PF's, i.e. of the gauged $SL(3)/U(1)\otimes U(1) $ - WZW model
and although it has rather complicated form, as explained in Sect.6 its soliton solutions(for $b_1=0$) 
can be easily obtained from the ones of the
 completely gauged IM(\ref{1.1}).



\end{document}